\documentstyle[12pt,epsfig,psfrag,amsmath]{article}

\newcommand {\f}{\frac}
\newcommand {\non}{\nonumber \\}
\newcommand {\bel}{\begin{equation}\label}
\newcommand {\beq}{\begin{equation}}
\newcommand {\eeq}{\end{equation}}
\newcommand {\beqa}{\begin{eqnarray}}
\newcommand {\beqal}{\begin{eqnarray}\label}
\newcommand {\eeqa}{\end{eqnarray}}
\newcommand {\bc}{\begin{center}}
\newcommand {\ec}{\end{center}}
\newcommand {\s}{\sigma}
\newcommand {\al}{\alpha^{'}}
\newcommand {\als}{\alpha^{'2}}
\newcommand {\th}{\theta}
\newcommand {\vth}{\vartheta}
\newcommand {\ka}{\kappa}
\newcommand {\pa}{\partial}
\newcommand {\de}{\delta}
\newcommand {\De}{\Delta}
\newcommand {\ep}{\epsilon}
\newcommand {\w}{\wedge}
\newcommand {\kpa}{k_{\parallel}}
\newcommand {\kpai}{k_{\parallel i}}
\newcommand {\kpaj}{k_{\parallel j}}
\newcommand {\kpe}{k_{\perp}}
\newcommand {\Tr}{\mbox{Tr}}

\newcommand {\ket}[1]{\left | #1 \right\rangle}
\newcommand {\expect}[1]{\left\langle #1 \right\rangle}
\newcommand {\expt}[1]{\left\langle #1 \right\rangle}
\newcommand {\alp}{\alpha}
\newcommand {\noi}{\noindent}
\newcommand {\ps}{\psi}
\newcommand {\bps}{\bar{\psi}}

\newcommand{\alpb}
{\left[\begin{smallmatrix}
\alp \\ \beta
\end{smallmatrix}\right]}

\newcommand{\oo}
{\left[\begin{smallmatrix}
1/2 \\ 1/2
\end{smallmatrix}\right]}

\newcommand{\zo}
{\left[\begin{smallmatrix}
0\\ 1/2
\end{smallmatrix}\right]}

\newcommand{\oz}
{\left[\begin{smallmatrix}
1/2 \\ 0
\end{smallmatrix}\right]}

\newcommand{\zz}
{\left[\begin{smallmatrix}
0 \\ 0
\end{smallmatrix}\right]}

\def\nn{\nonumber}

\def\vs5{\vspace*{5mm}}
\def\vs1{\vspace*{1cm}}
\def\vs2{\vspace*{2cm}}
\def\hs5{\vspace*{5mm}}
\def\hs1{\hspace*{1cm}}

\makeatletter

\@addtoreset{equation}{section}
\makeatother

\begin{document}

\title{
\hfill\parbox{4cm}{\normalsize IP/BBSR/2006-11}\\
\vspace{1cm}
UV/IR Mixing in Noncommutative Field Theories and Open Closed String 
Duality
\author{Swarnendu Sarkar \footnote{email:swarnen@iopb.res.in}\\
\\
{\em Institute of
Physics, Bhubaneswar}\\
{\em India 751 005}}}
\maketitle

\begin{abstract}

In this paper we study the phenomenon of UV/IR mixing in noncommutative 
field theories from the point of view of world-sheet open-closed 
duality in string 
theory. New infrared divergences in noncommutative field theories 
arise as a result of integrating over high momentum modes in the 
loops. These are believed to come from integrating out additional bulk 
closed string modes. We analyse this issue in detail for the bosonic 
theory and further for the supersymmetric theory on the $C^2/Z_2$ 
orbifold. We elucidate on the exact role played by the constant background 
$B$-field in this correspondence.

\end{abstract}
\newpage

\tableofcontents

\section{Introduction}

Dualities have played a very important role in the understanding of 
various issues in string theory. World-sheet duality in string 
theory have been observed since the early days of its formulation.
It is now believed that this duality underlies the duality between the 
open and closed string theories. In certain backgrounds such as in 
AdS/CFT \cite{adscft,nadscft}, this manifests as the relation between only the 
massless open and closed string sectors. In this review we will study 
this world-sheet duality in a background antisymmetric two-form 
constant $B$-field.

Studies of open string dynamics in the background $B$-field have shown 
that the low energy dynamics is given by a gauge theory on 
noncommutative space-times \cite{douglas1,schomerus,Ardalan:1998ks,chu,Seiberg}. Apart 
from this realisation of noncommutative 
gauge theory, quantum field theories on
noncommutative space-times have independently been studied for a long
time with a hope to cure the ultraviolet divergence problem in quantum 
field theories that
arise in the continuum limit. The
rationale for this being that discreteness of space time is inherent in
any quantum formulation of gravity and that noncommutative space-time
is one of the ways to achieve this 
\cite{Snyder:1946qz}\cite{Balachandran:1994pk}. Quantum 
field theories on
noncommutative backgrounds are nonlocal and sometimes violate
the conventional notions of local quantum field theories. Various
aspects of these theories have been studied extensively over the past
few years \cite{douglas}. More often their embedding into string 
theory has led to a better understanding of these issues. 

One of the
well known generic features of these theories is the mixing of the
ultraviolet and the infrared sectors contrary to the ordinary
quantum field theories where they decouple
\cite{minwalla}. Within the
domain of quantum field theory it is thus important to see how the usual 
notions of Wilsonian Renormalisation Group fits into these models. A 
thorough
analysis shows that an IR cut-off is necessary for the Wilsonian RG to
make sense here, and with the IR cut-off usual renormalisation can be
done \cite{myuvir}. The need for an IR cut-off in studying nonplanar 
anomalies in gauge theories has also been explored. See \cite{Ardalan:2005mj} 
and references therein.

Recently a different approach has been pursued that lead 
to the removal of UV/IR mixing in noncommutative field 
theories \cite{sachin,Balachandran:2005pn}. The spin-statistics theorem 
however needs to be modified here and thus it is,
not clear whether this is to be viewed as an inequivalent
quantisation and therefore a different theory or a different cure
to the infrared divergences (also see \cite{Chakraborty:2006hv}). 
For other approaches addressing this issue
of UV/IR mixing and hence renormalisation in noncommutative field theories
see \cite{Grosse:2004yu}.

In this review we will study this phenomenon of mixing of UV and the IR 
sectors of noncommutative field theories from the point of view of 
world-sheet open closed string duality. The ultraviolet divergences 
in open string theory can be interpreted as 
closed string infrared divergence using the world-sheet duality.
In the presence of the background $B$-field these divergences are 
regulated and thus a quantitative analysis can be made. 
The one-loop two point diagram for open strings is a cylinder with 
a modular parameter $t$ and vertex operator insertions at the 
boundaries. The two point one-loop noncommutative field theory diagram 
results in the Seiberg-Witten limit by keeping surviving terms in the 
integrand for the integral over $t$ for $t \rightarrow \infty$. This 
limit suppresses all contributions from massive modes in the loop. The 
resulting diagram is that of the gauge theory with massless propagating 
modes. This amplitude is usually divergent in the ultraviolet when 
integrated over $t$. The 
source of ultraviolet divergence is the same as that of those in string 
theory i.e. $t\rightarrow 0$. It is therefore natural to analyse the 
amplitude directly in this limit when only the low lying closed string 
exchanges contribute.

In
the bosonic string theory setting, we will first analyse the two-point 
one loop amplitude for gauge bosons on the brane, in the closed string
channel \cite{myopcl1}. 
Though there are additional
tachyonic divergences, we are able to show that the form of IR
divergences with appropriate tensor structures can be extracted by
considering only lowest lying modes (tachyonic and massless). We 
will further analyse the two point amplitude by studying massless closed 
string exchanges in background constant $B$-field. 

It was observed that ultraviolet behaviour of the one loop gauge theory 
is same as that of the infrared due to massless closed string tree-level 
exchanges in nonconformal gauge theory with ${\cal N}=2$ supersymmetry 
\cite{douglas2}. The simplest way to realise this gauge theory 
is on fractional branes localised at the fixed point on $C^2/Z_2$.
Various aspects of this duality have since been studied 
\cite{Klebanov:1999rd,lerda1,Polchinski:2000mx,lerda2,Bertolini:2001gq,Bertolini:2003,DiVecchia,DiVecchia:2005vm}.
In the next part of this analysis we will show that the 
UV/IR mixing phenomenon of 
${\cal N}=2$ gauge theory can be naturally interpreted as a consequence 
of open-closed string duality in the presence of background $B$-field 
\cite{myopcl2,myopcl3}.

This review is organised as follows. In Section~\ref{sbnc} we give a 
brief review 
of open bosonic strings in background constant $B$-field and the 
appearance of noncommutative field theory as low-energy description of 
$D$-brane world volume dynamics. In Section~\ref{osol}, we study the one 
loop 
open string amplitude in the UV limit and write down the contribution 
from the lowest
states. In Section~\ref{cse}, we analyse massless closed string exchanges 
in background $B$-field and reconstruct the massless contribution 
computed in Section~\ref{osol}. In Section~\ref{osbf} we study 
superstrings in $B$-field background 
and give a short review of strings on 
$C^2/Z_2$ orbifold and the massless spectrum of open strings ending on 
fractional $D_3$-brane localised at the fixed point and closed strings. 
In Section~\ref{tpa} we 
compute the two point function for one loop open strings in this 
orbifold background with the $B$-field turned on, and analyse it 
in the open and closed string channels. By taking the field theory 
limit, we show using open-closed string 
duality that the new IR divergent term from the nonplanar amplitude is 
exactly equal to the IR divergent contributions from massless closed 
string exchanges. In Section~\ref{ocse}, we study massless closed string exchanges on $C^2/Z_2$ orbifold along the same lines as in Section~\ref{cse}. We conclude this article with discussions in Section~\ref{dis}.\\
\\
\noindent
{\it Conventions:} We will use capital letters $(M, N,...)$ to denote
general space-time indices and small letters $(i,j, ...)$ for coordinates
along the $D$-brane. Small Greek letters $(\alpha,\beta...)$ will be used
to denote indices for directions transverse to the brane.

\section{Bosonic strings in background $B$-field and 
noncommutative field theory}\label{sbnc}

In this section we give a short review of open string dynamics in the 
presence of constant background $B$-field leading to noncommutative 
field theory on the world volume of a $D$-brane \cite{Seiberg}. In the 
presence of a constant background $B$-field, the world sheet action is 
given by,

\beqal{sb}
S_b=\f{1}{4\pi\al} \int_{\Sigma} [g_{MN}\pa_{a}X^{M}\pa^{a}X^{N} -2\pi 
i\al 
B_{MN}\epsilon^{ab}\pa_{a}X^{M}\pa_{b}X^{N}]
\eeqa

\noindent
Consider a $D_p$ brane extending in the directions $1$ to $p$, such 
that, $B_{MN} \neq 0$ only for $M,N \leq p+1$ and $B_{MN}=0$ for $M \leq 
p+1, N > p$. The equation of motion gives the following boundary 
condition,

\beqa
g_{MN}\pa_{n}X^{N}+2\pi i \al B_{MN}\pa_{t} X^{N}|_{\pa\Sigma}=0
\eeqa

The world sheet propagator on the boundary of a disc satisfying this 
boundary condition is given by,

\beqal{boundprop}
{\cal G}(y,y^{'})=-\al 
G^{MN}\ln(y-y^{'})^2+\f{i}{2}\th^{MN}\ep(y-y^{'})
\eeqa

\noindent
where, $\ep(\Delta y)$ is $1$ for $\Delta y>0$ and $-1$ for 
$\Delta y<0$. $G_{MN}$, $\th_{MN}$ are given by,

\beqal{gth}
G^{MN}&=&\left(\f{1}{g+2\pi\al B}g\f{1}{g-2\pi\al B}\right)^{MN}\non
G_{MN}&=&g_{MN}-(2\pi\al)^2(Bg^{-1}B)_{MN}\non
\th^{MN}&=&-(2\pi\al)^2\left(\f{1}{g+2\pi\al B}B\f{1}{g-2\pi\al 
B}\right)^{MN}
\eeqa

The relations above define the open string metric $G$ in terms of the 
closed string metric $g$ and $B$. This difference in the two metrics as 
seen 
by the open strings on the brane and the closed strings in the bulk 
plays an important role in the discussions in the following sections.
We next turn to to the low energy limit, $\al \rightarrow 0$. A 
nontrivial low energy theory results from the following scaling.

\beqal{swl}
\al \sim \epsilon^{1/2} \rightarrow 0 
\mbox{\hspace{0.1in};\hspace{0.1in}} 
g_{ij} \sim \epsilon \rightarrow 0
\eeqa

where, $i,j$ are the directions along the brane. This is the 
Seiberg-Witten (SW) limit that gives rise to noncommutative field 
theory on the brane. The relations in eqn(\ref{gth}), to the leading 
orders, in this 
limit reduce to,

\beqa
G^{ij}&=&-\f{1}{(2\pi\al)^2}(\th g \th)^{ij} 
\mbox{\hspace{0.1in};\hspace{0.1in}}
G_{ij}=-(2\pi\al)^2(Bg^{-1}B)_{ij}\non 
\th^{ij}&=&\left(\f{1}{B}\right)^{ij}
\eeqa    

for directions along the $D_p$ brane. $G_{MN}=g_{MN}$ and $\th=0$ 
otherwise. It was shown that the tree-level action for the low energy 
effective field theory on the brane has the following form,

\beqal{ncym}
S_{YM}=-\f{1}{g_{YM}^2}\int \sqrt{G} 
G^{kk^{'}}G^{ll^{'}}\Tr(\hat{F}_{kl}*\hat{F}_{k^{'}l^{'}})
\eeqa

\noindent
where the $*$-product is defined by, 

\beqa
f*g(x)=e^{\f{i}{2}\th^{ij}\pa_{i}^y\pa_{j}^z}f(y)g(z)\mid_{y=z=x}
\eeqa

\noindent
and $\hat{F}_{kl}$ is the noncommutative field strength, which is 
related to the ordinary field strength, $F_{kl}$ by the 
Seiberg-Witten map,

\beqal{redeff}
\hat{F}_{kl}=F_{kl}+\th^{ij}(F_{ki}F_{lj}-A_i\pa_{j}F_{kl})+{\cal 
O}(F^3)
\eeqa

\noindent
and,

\beqa
\hat{F}_{kl}=\pa_k\hat{A}_l-\pa_l\hat{A}_k-i\hat{A}_k*\hat{A}_l+
i\hat{A}_l*\hat{A}_k
\eeqa

This form of the tree-level action (\ref{ncym}) is derived from the 
n-point
tree-level open string correlators with gauge field vertices and then
keeping the surviving terms in the low energy limit (\ref{swl}). The
vertex is given by (\ref{photonvertex}) and the boundary propagator is
(\ref{boundprop}). One of the most important features of 
these noncommutative field theories is the coupling of 
the UV and the IR regimes, manifested by the nonplanar sector 
of these theories, contradicting our usual notions of Wilsonian 
RG \cite{minwalla}. To see this, let us consider a noncommutative scalar 
($\lambda\phi^4$) theory in four dimensions. The noncommutative theory 
is written with all the 
products of fields replaced by $*$-products. The nonplanar 
one-loop two-point amplitude has the following form, 

\beqal{scalar2pt}
\Gamma^2_{NP}(p)\sim 
\Lambda^2_{eff}-m^2\ln\left(\f{\Lambda^2_{eff}}{m^2}\right)
\eeqa

\noindent
where, $\Lambda^2_{eff}=1/(1/\Lambda^2+pop)$, $pop=-(\th p)^2$ and 
$\Lambda$ is the UV cut-off. The amplitude is finite in the UV but is IR 
divergent, though we had a massive theory to start with. 
Note that $(\th p)^2$ plays the role of $1/\Lambda^2$ in the 
continuum limit. It 
was suggested \cite{minwalla} that these IR divergent terms could 
arise 
by integrating out massless modes at high energies. The effective action 
containing the two point function can be written as,

\beqa
S^{'}=S(\Lambda)+\int d^4x \left[\f{1}{2}\pa \chi o \pa \chi +\f{1}{2}
\Lambda^2\left(\pa o 
\pa\chi\right)^2+i\f{1}{\sqrt{96\pi^2}}\lambda\chi\phi\right]
\eeqa

$S(\Lambda)$ is the effective action for the cut-off field theory and 
$\chi$ is a massless field. Integrating out $\chi$ gives the 
quadratic piece in the effective action of the original theory in the 
continuum limit. It was further noted 
that both the quadratic and the log terms of eqn(\ref{scalar2pt}) can 
be recovered through massless 
tree-level exchanges if these modes are allowed to propagate in $0$ and 
$2$ extra dimensions transverse to the brane respectively 
\cite{raamsdonk}. This is quite
like the open string one loop divergence which is reinterpreted as IR
divergence coming from massless closed string exchange.

A similar 
structure arises for the nonplanar two point function for the gauge 
boson in noncommutative gauge theories,

\beqal{photon2pt}
\Pi^{ij}(p) \sim 
N_1[G^{ij}G^{kl}-G^{ik}G^{jl}]p_kp_l\ln(p^2\tilde{p}^2) 
+N_2 \f{\tilde{p}^{i}\tilde{p}^{j}}{\tilde{p}^4}
\eeqa

\noindent
$N_1$ and $N_2$ depends on the matter content of the theory. For some 
early works on noncommutative gauge theories see \cite{martin}.
The effective action with the two point function (\ref{photon2pt}) is 
not gauge invariant.
To write down a gauge invariant effective action one needs to introduce 
open Wilson lines \cite{kawai} 

\beqa
W_C(p)=\int d^4x P*\exp\left(ig\int_C d\s \pa_{\s}y^{i} 
A_{i}(x+y(\s)\right)*e^{ipx}
\eeqa

\noindent
The curve $C$ is parametrised by $y^{i}(\s)$, where $0\leq \s \leq 1$ 
such that, $y^{i}(1)-y^{i}(0)=\tilde{p}^{i}$. 
Correlators of Wilson lines in noncommutative gauge theories have been 
studied by various authors \cite{wilsonline}. The terms in 
(\ref{photon2pt}) are the leading terms in the expansion of 
the two point function for the open Wilson line.
A crucial point to be noted is that for supersymmetric theories, $N_2$, 
the coefficient of the second term, which is allowed by the 
noncommutative gauge invariance vanishes \cite{matusis}. 
Also see \cite{khoze} for an elaborate discussion. An 
observation on the arising of tachyon in the closed string theory 
in the bulk and the non vanishing of $N_2$ with a negative sign was made 
in \cite{armoni}. Thus when the closed string theory is unstable due 
to the presence of tachyons, the two point 
function in noncommutative gauge theory also diverges with a negative 
sign for low momenta.  
Various attempts have 
been made, along the lines as discussed above, to recover the nonplanar
IR divergent terms from tree-level closed string exchanges 
\cite{oneloop1,oneloop2,oneloop4,oneloop5,rajaraman,Chaudhuri}. 
This is the 
issue that we shall address in the following 
sections.

\section{Open string one loop amplitude : Bosonic theory}\label{osol}

In the previous section we have outlined how new IR divergent terms 
appear in the nonplanar loop amplitudes of noncommutative field 
theories. With a view to interpret these in terms of closed string 
exchanges we now embed the problem in string theory. The main idea is 
summarised in Figure~\ref{lim2}. The various steps involved in the 
problem will be clarified as we proceed.
In this section we compute the open string one loop amplitude with 
insertion of two gauge field vertices. We will compute the two point 
amplitude in the closed string channel keeping only the contributions 
from the tachyon and the massless modes. One loop amplitudes for open 
strings with two vertex insertions in the presence of a constant 
background $B$-field have been computed by various 
authors, and field theory amplitudes were obtained in the 
$\al\rightarrow 0$ limit 
\cite{oneloop1,oneloop2,oneloop3,oneloop4,oneloop5,Chaudhuri}. 

\noindent
The one loop partition function is written as 
\cite{callan,polchinski}

\beqa
Z(t)=\det(g+2\pi\al B){\cal V}_{p+1}(8\pi^2\al
t)^{-\f{p+1}{2}}Z_0(t)
\eeqa

with,
\beqa
Z_0(t)= \Tr[\exp(-2\pi t L_0^{'})]
\eeqa

\noindent
$\det(g+2\pi\al B)$ comes from the trace over the zero modes of the 
world-sheet bosons. See Appendix~\ref{vacuum} eqn(\ref{det}), $t$ is 
the modulus of the cylinder and $L_0^{'}$ contains the oscillators. This 
gives,

\beqal{zt}
Z(t)=\det(g+2\pi\al B){\cal V}_{p+1}(8\pi^2\al 
t)^{-\f{p+1}{2}}\eta(it)^{-(D-2)}
\eeqa

\noindent
${\cal V}_{p+1}$ is the volume of the $D_p$ brane.
We are interested in the non-planar two point one loop amplitude that is
obtained by inserting the two vertices at the two different boundaries on 
the cylinder. 

\beqa
A(p_1,p_2)=\int_{0}^{\infty}\f{dt}{2t}Z(t)\int_{0}^{2\pi 
t}dy\int_{0}^{2\pi t}dy^{'} 
<V(p_1,x,y)V(p_2,x^{'},y^{'})>
\eeqa

\noindent
where $Z(t)$ is as defined in eqn(\ref{zt}). The required vertex 
operator is given 
by, 

\beqal{photonvertex}
V(p,y)=-i\f{g_o}{(2\al)^{1/2}}\ep_{j}\pa_y 
X^{j}e^{ip.X}(y)
\eeqa

\noindent
The noncommutative field theory results are recovered from region of the 
modulus where $t\rightarrow \infty$ in the SW limit. As mentioned, the 
nonplanar diagrams in the noncommutative field theory give rise to 
terms which manifest coupling of the UV to the IR sector of the 
field theory. 

The $t \rightarrow 0$ limit, picks out the contributions only 
from the tree-level massless closed string exchange. This is the UV 
limit of the open string. The amplitude is usually divergent. However, 
in the usual case, these divergences are reinterpreted as IR divergences 
due to the 
massless closed string modes. What is the role played by the $B$-field? 
In the presence of the background $B$-field, the integral over the 
modulus is regulated. On the closed string side, this would mean that 
the propagator for the massless modes are modified. In the following parts
we will analyse this region of the modulus, $t$.

\begin{figure}[t]
\begin{center}
\begin{psfrags}
\psfrag{i}[][]{$\infty$}
\psfrag{z}[][]{$0$}
\psfrag{l}[][]{$\Lambda$}
\psfrag{Open}[][]{Open String Channel}
\psfrag{ls}[][]{$1/\Lambda^2\al$}
\psfrag{uv}[][]{UV}
\psfrag{p}[][]{$k$}
\psfrag{t}[][]{$t$}
\psfrag{a}[][]{(i)}
\psfrag{b}[][]{(ii)}
\psfrag{c}[][]{(iii)}
\psfrag{d}[][]{(iv)}
\psfrag{la}[][]{$1/\Lambda\al$}
\psfrag{las}[][]{$\Lambda^2\al$}
\psfrag{Closed}[][]{Closed String Channel}
\psfrag{ir}[][]{IR}
\psfrag{kpe}[][]{$\kpe$}
\psfrag{s}[][]{$s$}
\epsfig{file=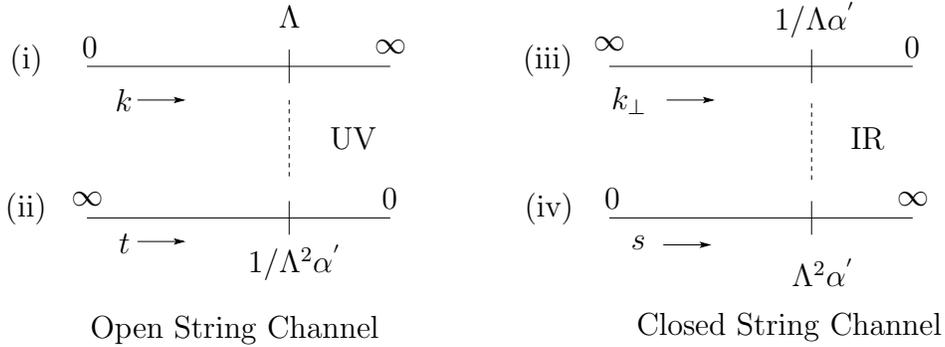, width= 12cm,angle=0}
\end{psfrags}
\vspace{ .1 in }
\caption{UV and IR regions in the open and closed string channels. (i) 
$k$ is the momentum in the gauge theory one loop diagram. (ii) $t$ is 
the modulus of the cylinder in the open string one loop diagram (iii) 
$\kpe$ is the transverse momentum of the closed string modes emitted 
from the brane (iv) $s=1/t$}
\label{lim1}
\end{center}
\end{figure}

Before going into the actual form let us see heuristically what 
we can expect to compare on both ends of the modulus.
First consider the one loop amplitude,

\beqal{op}
{\cal A} \sim \int \f{dt}{t} (\al t)^{-\f{p+1}{2}} \eta(it)^{-(D-2)} 
\exp(-C/\al t)
\eeqa

where $C$ is some constant depending on the 
$B$-field. In the $t \rightarrow \infty$ limit,

\beqal{opc}
{\cal A}_{op} \sim \int \f{dt}{t} (\al t)^{-\f{p+1}{2}} 
\left[e^{2\pi t}+(D-2)+O(e^{-2\pi t})\right]\exp(-C/\al t)
\eeqa

If we remove the tachyon, and restrict ourselves only to the $O(1)$ 
term in the expansion of the $\eta$-function, we see that 
$\al$ and $t$ occur together. 
This means that in the $\al \rightarrow 0 $ limit the finite 
contributions to the field theory come from the region where $t$ is 
large. We can break the integral over $t$ into two parts, 
$1/\Lambda^2\al<t<\infty$ and $0<t<1/\Lambda^2\al$, where $\Lambda$ 
translates into the UV cut-off for the field theory on the brane. The 
second interval is the source of divergences in the field 
theory that is regulated by $C$. This is the region of 
the modulus dominated by massless exchanges in the closed string 
channel. See Figure~\ref{lim1}. For the closed string channel, we have

\beqal{clc}
{\cal A}_{cl} \sim \int ds (\al)^{-\f{p+1}{2}}s^{-l/2}
\left[e^{2\pi s}+(D-2)+O(e^{-2\pi s})\right]\exp(-Cs/\al )
\eeqa

where $l=D-(p+1)$, is the number of dimensions transverse to the $D_p$ 
brane. The divergences as $C \rightarrow 0$ are regulated as 
as $1/C$ or $\ln(C)$, depending on $l=0$ or $l=2$ respectively 
\cite{raamsdonk}. The full open 
string channel result, including the finite contributions, 
will always require all the closed string modes 
for its dual description. As far as the divergent (UV/IR mixing) terms are 
concerned, we can hope to realise them through some field theory of the 
massless closed string modes. However, the exact correspondence between the 
divergences in both the channels, is destroyed by the presence of the 
tachyons. Also note that, at the $t \rightarrow 0$ end of 
the open string one loop amplitude, the divergence 
is contributed by the full
tower of open string modes. However, in the cases where the one loop open
string amplitude restricted to only the
massless exchanges can be rewritten as massless closed string 
exchanges, the integrand as a function of $t$ in one loop
amplitude should have the same asymptotic form as $t\rightarrow 0$ and
$t \rightarrow \infty$ so that eqn(\ref{clc}) is exactly the same as 
that of eqn(\ref{opc}) 
integrated between $[0,1/\Lambda^2\al]$. 
Examples of such configurations occur in some supersymmetric theories
where the one
loop open string amplitude restricted to the massless sector can be
rewritten exactly as tree-level massless closed string exchanges. It was 
shown that in these situations the potential between two branes
with separation $r$ is the same at both the $r\rightarrow 0$, and
$r\rightarrow \infty$ corresponding to
$t\rightarrow \infty$ and $t\rightarrow 0$ ends
respectively \cite{douglas2}. 
This has lead to further interesting studies on the gauge/gravity 
correspondence. For a review see 
\cite{DiVecchia:2005vm} and 
references therein. We can
expect that in these cases the IR singularities of the noncommutative
gauge theory match with those computed from the closed string massless
exchanges. However in the following analysis of the bosonic theory we will 
set $l=2$ so as to reproduce UV/IR effects of the 4D noncommutative gauge 
theory. We will see that this exercise will be helpful in uncovering various
details involved in the correspondence.

Let us now return to the original computation of the amplitude in the closed
string channel. The nonplanar world sheet propagator obtained by 
restricting to the 
positions at the two boundaries is,  

\beqa
{\cal G}^{ij}(y,y^{'})=-\al 
G^{ij}\ln\left|e^{-\f{\pi}{4t}}
\f{\vth_4\left(\f{\Delta y}{2\pi t},\f{i}{t}\right)}
{t^{-1}\eta(i/t)^3}\right|^2
-i\f{\th^{ij}\Delta y}{2\pi t}
-\al g^{ij}\f{\pi}{2t}
\eeqa

\noindent
where, $\Delta y=y-y^{'}$.
In the limit $t\rightarrow 0$ the propagator has the following 
structure,
  
\beqa
{\cal G}^{ij}=-4\al G^{ij}
\left[cos(\Delta y/t)e^{-\f{\pi}{t}}-e^{-\f{2\pi}{t}}\right]
-i\f{\th^{ij}\Delta y}{2\pi t}
-\al g^{ij}\f{\pi}{2t}
\eeqa

\noindent
Inserting this into the correlator for two gauge bosons and keeping 
only terms that would contribute to the tachyonic and massless closed 
string exchanges, we get,

\beqa
A_2(p,-p)=-i\det(g+2\pi\al B){\cal V}_{p+1}(\f{g_o^2}{2\al})
(8\pi^2\al )^{-\f{p+1}{2}}\ep_{i}\ep_{j}I(p)
\eeqa

\noindent
with $I(p)=I_{T}(p)+I_{\chi}(p)$ and,

\def\note{Note that the $s$ integral has to be cut off at the lower end 
at some value 
$\Lambda^2\al$. This corresponds to the UV transverse momentum cut-off 
for the closed strings, that allows us to extract the contribution from 
the IR region (see Figure~\ref{lim1}).
\beqal{cut-off}
I(p,\Lambda)\sim 
\int_{\Lambda^2\al}^{\infty}\f{ds}{s}e^{-p^2\al s}
\sim \int_{0}^{\infty}d^2\kpe 
\f{e^{-(\kpe^2+p^2)\Lambda^2\als}}{(\kpe^2+p^2)\al}
\eeqa
The integral over $\kpe$, eqn(\ref{cut-off}) receives contribution upto 
$\kpe\sim 
O(1/\Lambda\al)$.
The included region of the $\kpe$ integral is the required IR sector for 
the transverse closed string modes or the UV for the open string 
channel.
}

\beqal{T}
I_{T}(p)&=&\tilde{p}^{i}\tilde{p}^{j}\int 
ds s^{-\f{l}{2}}\exp\left\{-(\f{\al\pi}{2}p_{i}g^{ij}p_{j}
-2\pi)s\right\}\\
&=&4\pi(2\pi^2\al)^{\f{l}{2}-1}\tilde{p}^{i}\tilde{p}^{j}\int 
\f{d^l\kpe}{(2\pi)^l}\f{1}{\kpe^2+p_{i}g^{ij}p_{j}-4/\al}
\nonumber
\eeqa

In (\ref{T}) we have written the integral over $t$ in terms of $s=1/t$. 
Further in the last expression we have replaced the integral over 
$s$ with that of $\kpe$.
The dimension of the $\kpe$ integral, $l$ is the number of directions 
transverse to the brane and is thus the momentum of the closed string along 
these directions.
\note{}. With this observation, for the tachyon with $l=2$, we get

\beqal{tac2pt}
I_{T}(p,\Lambda)&=&4\pi^2(2\pi^2\al)^{\f{l}{2}-1}\tilde{p}^{i}\tilde{p}^{j}
\ln\left(\f{p_{i}g^{ij}p_{j}-4/\al+\f{1}{(\Lambda\al)^2}}
{p_{i}g^{ij}p_{j}-4/\al}\right)
\eeqa

\noindent
For the noncommutative limit (\ref{swl}), we can expand the answer 
(\ref{tac2pt}) in powers of 
$1/(\al pg^{-1}p)$,

\beqal{texp}
\ln\left(pg^{-1}p-4/\al\right)\sim 
\ln\left(pg^{-1}p\right)-\f{4}{\al pg^{-1}p}-
\f{1}{2}\left(\f{4}{\al pg^{-1}p}\right)^2 - \textellipsis
\eeqa

\noindent
The $(1/\al pg^{-1}p)^2$ term in the expansion 
(\ref{texp}) above  
corresponds to the IR singular term which appears in the noncommutative 
gauge theory. To compare with the second term of
(\ref{photon2pt}), we should set $G=\eta$, so that $g^{-1} \sim 
-\th^2/\als $. As far as the exact coefficient is 
concerned, the full tower of massive states would contribute. 
The absence of this term in the supersymmetric theories can only be due 
to exact cancellations between the bosonic and fermionic sector 
contributions \cite{armoni}.

Similarly we now write down the contribution from 
the massless exchanges,

\beqal{clchannel}
I_{\chi}(p,\Lambda)&=&4\pi(2\pi^2\al)^{\f{l}{2}-1}\left[(D-2)
\tilde{p}^{i}\tilde{p}^{j}+8(2\pi\al)^2 p_{k}p_{l}
(G^{ij}G^{kl}-G^{ik}G^{jl})\right]\times\non
&\times&
\int
\f{d^l\kpe}{(2\pi)^l}\f{1}{\kpe^2+p_{i}g^{ij}p_{j}}
\eeqa

\noindent
One can observe that the terms occurring with $\als (\sim \ep)$ as the 
coefficient, relative to the other terms in
(\ref{texp}) and (\ref{clchannel}), appear 
in the gauge theory result in eqn(\ref{photon2pt}).
In the closed string channel we have got this for the number of 
transverse dimensions, $l=2$. This means that $p+1=D-2=24$ is the 
dimension of the gauge theory on the string side. However the result of 
eqn(\ref{photon2pt}) is valid for the noncommutative gauge theory 
defined in 
4-dimensions.
To understand why it is these terms that occur in the four dimensional 
gauge theory, we must have a string setting where $l=2$ and $p=3$.
At this point, as discussed earlier, it is only necessary that 
$l=2$ so 
that the lowest lying closed string exchanges reproduce the
correct form of the IR singularities as that of the gauge theory in 
eqn(\ref{photon2pt}).

The exact correspondence between the UV behaviour 
of the noncommutative gauge theory and closed string exchanges 
would require the full tower of closed string states in this bosonic 
case. The contribution 
from the massive closed string states are likely to be suppressed only in 
some supersymmetric configurations 
\cite{douglas2}-\cite{DiVecchia:2005vm}. 
We will see how this works out in these setups in Sections~\ref{osbf}, 
\ref{tpa} and \ref{ocse}, 
but before that let us study the massless closed string 
exchanges in the presence of background $B$-field.


\section{Massless closed string exchanges}\label{cse}

In this section we reconstruct the two point function of two gauge 
fields eqn(\ref{clchannel})
from massless closed string exchanges. The aim here is to write the 
amplitude as sum of massless closed string exchanges in the presence of
constant background $B$-field. By considering the effective field theory 
of massless
closed strings, we construct the propagators for these modes 
(graviton, dilaton and the antisymmetric two-form field) with a constant 
background $B$-field. We compute the couplings of 
the gauge field on the brane with the massless closed strings from the 
DBI action. Finally we combine these results to write down the two point 
function. We will consider three separate cases when computing the two 
point amplitude in this section. 

\begin{enumerate}

\item In this case the background $B$-field 
is assumed to be small and the closed string metric, $g=\eta$.
The amplitude will be analysed to ${\cal O}(B^2)$.

\item The Seiberg Witten limit, when $g=\ep\eta$ with the amplitude expanded to
 ${\cal O}(\ep^2/(2\pi\al)^2)$.

\item The case when the open
string metric on the brane, $G=\eta$ so that $g=-(2\pi\al)^2B^2 +{\cal 
O}(\alpha^{'4})$ and the amplitude will be expanded to
${\cal O}((2\pi\al)^2)$

\end{enumerate}

The amplitude eqn(\ref{clchannel}) in the closed 
string channel is the closed form result of the massless exchanges. In 
each of the above cases, we will compare this amplitude to respective
orders with the ones we compute here in this section.
Let us begin by considering the field theory of the massless modes 
of the closed string string propagating in the bulk. The space-time action for 
closed string fields is written as,

\beqa
S=\f{1}{2\ka^2}\int d^DX
\sqrt{-g}[R-\f{1}{12}
e^{-\f{8\phi}{D-2}}H_{LMN}H^{LMN}
-\f{4}{D-2}g^{MN}\pa_{M}\phi\pa_{N}\phi]
\eeqa

where $D$ is the number of dimensions in which the closed string
propagates. The indices are raised and lowered by $g$. We will now 
construct the tree-level propagators that will 
be necessary in the next section to compute two point amplitudes. For 
each of the cases as defined above, the propagator will take a 
different form. Let us first consider the dilaton. For $g=\eta$ the 
propagator is the usual one,

\beqal{dp1}
<\phi\phi>&=&-\f{(D-2)i\ka^2}{4}\f{1}{\kpe^2+\kpa^2}
\eeqa

The next limit for the metric is $g=\ep\eta$ along the world volume 
directions of the brane. In this limit, the dilaton 
part of the action can be written as,

\beqa
S_{\phi}=-\f{4}{\kappa^2(D-2)}\int d^DX
\f{1}{2}[\pa_{\alpha}\phi\pa^{\alpha}\phi+\ep^{-1}
\pa_{i}\phi\pa^{i}\phi]
\eeqa

This gives the propagator,

\beqal{dp2}
<\phi\phi>&=&-\f{(D-2)i\ka^2}{4}\f{1}{\kpe^2+\ep^{-1}\kpa^2}
\eeqa

Finally, when the open string metric is set to, $G=\eta$,
the lowest order solution for $g$ along the brane directions is,

\beqa
g=-(2\pi\al)^2B^2 +{\cal O}(\alpha^{'4})
\eeqa

which gives,

\beqal{dp3}
<\phi\phi>&=&-\f{(D-2)i\ka^2}{4}\f{1}{\kpe^2+\tilde{\kpa}^2/(2\pi\al)^2}
\eeqa

where,
\beqa
\tilde{\kpa}^2=-\kpai\left(\f{1}{B^2}\right)^{ij}\kpaj
\eeqa

Let us now turn to the free part for the antisymmetric tensor field,

\beqa
S_b=-\f{1}{24\kappa^2}\int d^DX H_{LMN}H^{LMN}
\eeqa

where,
\beqa
H_{LMN}=\pa_{L}b_{MN}+\pa_{M}b_{NL}
+\pa_{N}b_{LM}
\eeqa

Using the following gauge fixing condition,

\beqa
g^{MN}\pa_{M}b_{NL}=0
\eeqa

The action reduces to,

\beqa
S_b=-\f{(2\pi\al)^2}{8\kappa^2}\int d^DX 
\left[g^{\alpha\beta}\pa_{\alpha}
b_{IJ}\pa_{\beta}b_{KL}+g^{ij}\pa_i b_{IJ}\pa_j 
b_{KL}\right]g^{IK}g^{JL}
\eeqa

The factor of $(2\pi\al)^2$ in the $b$-field action has been 
included because
the sigma model is defined with $(2\pi\al)B$ coupling.
The propagator then is,

\beqal{bp}
<b_{IJ}b_{I^{'}J{'}}>=-\f{2i\ka^2}{(2\pi\al)^2}\f{g_{I[J{'}}g_{I^{'}]J}}
{\kpe^2+g^{ij}\kpai\kpaj}
\eeqa

Finally, the gravitational part of the action. As will turn out in the 
next section that we will only have to consider graviton exchanges for 
the case $g=\eta$. The propagator for the graviton here is the usual 
propagator from the action,

\beqa
S_h=\f{1}{2\ka^2}\int d^DX
\sqrt{-g}R
\eeqa

By considering fluctuations about $\eta$, and in the gauge 
(\ref{hgauge}),

\beqa
g_{MN}=\eta_{MN}+h_{MN} 
\eeqa

\beqal{hgauge}
g^{MN}\Gamma^{L}_{MN}=0
\eeqa

the graviton propagator is,

\beqal{gp}
<h_{IJ}h_{I^{'}J{'}}>=-2i\ka^2\f{[\eta_{I\{J{'}}\eta_{I^{'}\}J}-2/(D-2)
\eta_{IJ}\eta_{I^{'}J{'}}]}{\kpe^2+\kpa^2}
\eeqa

After writing down the required propagators, we now turn to the 
computation of the vertices. As mentioned in the beginning of this 
section, we will consider each of the three cases separately. To begin,
we first write down the DBI action for a $D_p$ brane,

\beqa
S_p=-T_p\int d^{p+1}\xi e^{-\Phi}\sqrt{g^{'}+2\pi\al(B+b)}
\eeqa

Where, $g^{'}$ is the closed string metric in the string frame, $B$ is 
the constant two form
background field and $b$ is the fluctuation of the two form field. The
$b$-field on the brane is interpreted as the two form field strength 
for
the $U(1)$ gauge field and in the bulk it is the usual two form 
potential. Going to the Einstein frame by defining,

\beqa
g =g^{'}e^{2\omega}\mbox{;\hspace{0.2in}}
\omega=\f{2(\phi_0-\Phi)}{D-2}\mbox{;\hspace{0.2in}}
\Phi=\phi+\phi_0 \mbox{;\hspace{0.2in}}
\omega=\f{-2\phi}{D-2}
\eeqa

the action can be rewritten as,

\beqal{bi}
S_p&=&-\tau_p\int d^{p+1}\xi {\cal L}(\phi,,h,b)\non
&=&-\tau_p\int d^{p+1}\xi e^{-\phi(1-\f{2(p+1)}{D-2})}
\sqrt{g+2\pi\al(B+b)e^{-\f{4\phi}{D-2}}}
\eeqa

where, $\tau_p=T_pe^{-\phi_0}$ and $\phi$ is the propagating dilaton
field. We will now consider each of the three cases separately and 
compute the two point function upto the respective orders.

\subsection{Expansion for small $B$}\label{cse1}

\begin{figure}[t]
\begin{center}
\begin{psfrags}
\psfrag{B}[][]{\footnotesize{$B$} }
\psfrag{F}[][]{\footnotesize{$F$} }
\psfrag{b}[][]{$b$ }
\psfrag{h}[][]{$h$ }
\psfrag{p}[][]{$\phi$ }
\psfrag{i}[][]{\footnotesize{(i)}}
\psfrag{ii}[][]{\footnotesize{(ii)}}
\psfrag{iii}[][]{\footnotesize{(iii)}}
\epsfig{file=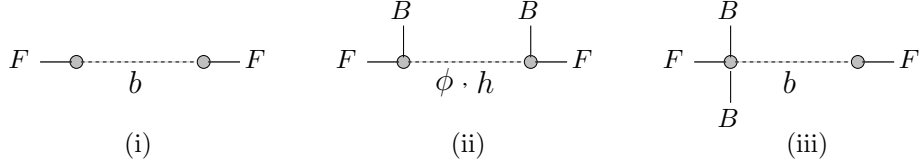, width= 12 cm,angle=0}
\end{psfrags}
\vspace{ .1 in }
\caption
{Two point amplitude upto quadratic order in $B$. (i) and (iii) are due 
to 
only 
$b$-field exchange, (ii) is due to graviton and dilaton exchange.}
\label{smallb}
\end{center}
\end{figure}

In this part we compute the couplings of the gauge field on the brane to 
the massless closed strings in the bulk. We will assume the background 
constant $B$-field to be small and compute the lowest order 
contribution to the two point function considered as an expansion in 
$B$. The first thing to note is that, since $B$ is antisymmetric,
there cannot be a non vanishing amplitude with a single $B$ in one vertex
only. We need at least two powers of $B$. 
One on each vertex or both on one. The graviton and the dilaton
need one on each vertex. The $b$-field  can couple to the gauge field
without a $B$. So for the $b$-field we need to consider couplings upto
${\cal O}(B^2)$. The closed string tree-level diagrams contributing to 
the three massless modes are shown in Figure~\ref{smallb}.

\beqa
{\cal L}=\sqrt{e^{-P\phi}\left[g+(2\pi\al)e^{-Q\phi}(B+b)\right]}
\eeqa

\noindent
where
\beqa
P=\f{2}{p+1}-\f{4}{(D-2)} \mbox{\hspace{0.2in}} Q=\f{4}{(D-2)}
\eeqa

The vertices can now be obtained by expanding ${\cal L}$ for small 
$B$, with $g=\eta+h$,

\beqal{exp}
{\cal L}=\sqrt{e^{-P\phi}[g+(2\pi\al)e^{-Q\phi}b]}
\left[1+\f{(2\pi\al)e^{-Q\phi}}{g+(2\pi\al)e^{-Q\phi}b}B\right]^{1/2}
\eeqa

The vertices for the graviton and dilaton and the $b$-field are,

\beqa
V_h^{ij}&=&-\tau_p(2\pi\al)^2\left[-\f{1}{4}B^{kl}\eta^{ij}
+\eta^{jl}B^{ki}\right]\\
V_{\phi}&=&-\tau_p(2\pi\al)^2\left[\f{1}{4}(p+1)P+Q\right]B^{kl}
\eeqa

\beqa
V_b^{ij}&=&\tau_p\f{(2\pi\al)^2}{2}\eta^{ki}\eta^{lj}
\left(1-(2\pi\al)^2\f{1}{4}\Tr(B^2)\right)
\non
&-&\tau_p(2\pi\al)^4\left[-\f{1}{4}B^{kl}B^{ij}-\f{1}{2}B^{ki}B^{lj}
-(B^2)^{li}\eta^{jk}\right]
\eeqa

The propagators are the usual ones, rewriting them from eqns(\ref{dp1},
\ref{bp},\ref{gp}),

\beqa
<h_{ij}h_{i^{'}j^{'}}>&=&-2i\ka^2\f{[\eta_{ii^{'}}\eta_{jj^{'}}+
\eta_{ij^{'}}\eta_{i^{'}j}-2/(D-2)\eta_{ij}\eta_{i^{'}j^{'}}]}{\kpe^2+\kpa^2}\\
<\phi\phi>&=&-\f{(D-2)i\ka^2}{4}\f{1}{\kpe^2+\kpa^2}\\
<b_{ij}b_{i^{'}j^{'}}>&=&-\f{2i\ka^2}{(2\pi\al)^2}\f{[\eta_{ii^{'}}\eta_{jj^{'}}-\eta_{ji^{'}}
\eta_{ij^{'}}]}{\kpe^2+\kpa^2}
\eeqa

\noindent
With these, the contributions from the three modes to the two point 
function can be worked out. 
We are interested in the correction to the quadratic term in the 
effective action for the 
gauge field on the brane. This can be constructed with the vertices 
computed above and the propagators for the intermediate massless closed 
string states. This correction for the nonplanar diagram can be written 
as, 

\beqal{pos2pt}
A_2(bb)=\int d^{p+1}\xi\int d^{p+1}\xi^{'}b(\xi)b(\xi^{'})
V_{\chi}<\chi(\xi)\chi(\xi^{'})>V_{\chi}
\eeqa
\noindent
where,
\beqa
<\chi(\xi)\chi(\xi^{'})>=\int 
\f{d^Dk}{(2\pi)^D}<\chi(\kpe,\kpa)\chi(-\kpe,-\kpa)>e^{-i\kpa(\xi-\xi^{'})}
\eeqa

\noindent
We can rewrite eqn(\ref{pos2pt}) in momentum space coordinates as,
 
\beqal{eff}
A_2(bb)&=&{\cal V}_{p+1}\int \f{d^{p+1}p}{(2\pi)^{p+1}}b(p)b(-p)\int 
\f{d^l\kpe}{(2\pi)^l}V_{\chi}<\chi(\kpe,-p)\chi(-\kpe,p)>V_{\chi}\non
&=&{\cal V}_{p+1}\int \f{d^{p+1}p}{(2\pi)^{p+1}}b(p)b(-p)L_2(p,-p)
\eeqa

\noindent
In the planar two point function, both the vertices are on the same end 
of the
cylinder in the world-sheet computation. In the field theory this 
corresponds to putting both the
vertices at the same position on the $D$-brane. In other words,
in the expansion of the DBI action, we should be looking for $b^2\chi$
vertices on one end and a $\chi$ tadpole on the
other. In this case, from the above calculation, $\kpa=0$. So 
the closed string propagator is just $1/\kpe^2$, i.e. the 
propagator is not modified by the momentum of the gauge field on the 
brane. This is what we expect, as in the field theory on the brane, the 
loop integrals are not modified for the planar diagrams. Here we will 
only concentrate on the nonplanar sector.
\\

\noindent
As mentioned earlier, on the brane we will identify, 
\beqal{ident}
b_{kl}(p)\equiv \f{g_0}{\sqrt{2\al}}F_{kl}(p)
=\f{g_0}{\sqrt{2\al}}p_{[k} A_{l]}(p)
\eeqa

\noi
The two point amplitude including the graviton, dilaton and $b$ field exchange is given by, 

\beqal{final1}
L_2&=&-i\ka^2\tau_p^2\int\f{d^l\kpe}{(2\pi)^l}
\f{1}{\kpe^2+p^2}\times\\&\times&[(2\pi\al)^4\f{D-2}{32}B^{kl}B^{k^{'}l^{'}}
+\f{(2\pi\al)^2}{4}\{1-\f{(2\pi\al)^2}{2}\Tr(B^2)\}(\eta^{ll^{'}}\eta^{kk^{'}}-
\eta^{lk^{'}}\eta^{kl^{'}})\non
&+&\f{(2\pi\al)^4}{2}\{(B^2)^{kk^{'}}\eta^{ll^{'}}-
(B^2)^{kl^{'}}\eta^{lk^{'}}\}+(kl)\leftrightarrow(k^{'}l^{'})]\nonumber
\eeqa

\noi
The full two point effective action, can now be constructed by putting 
back $L_2$ in eqn(\ref{eff}) along with the identification 
eqn(\ref{ident}).
To compare this with the closed string channel result with only massless 
exchanges, eqn(\ref{clchannel}) we must note the expansions of the 
following quantities to appropriate powers of $B$.

\beqal{expan1}
G^{ij}&\sim & \eta^{ij}+(2\pi\al)^2(B^2)^{ij}+{\cal O}(B^4)\\
\th^{ij}&\sim &-(2\pi\al)^2B^{ij}+{\cal O}(B^3)\non
\sqrt{\eta+(2\pi\al)B}&\sim &\left[1-\f{(2\pi\al)^2}{4}\Tr(B^2)
+{\cal O}(B^4)\right]\nonumber
\eeqa

With these expansions, we can see that eqn(\ref{clchannel}) equals the 
sum of massless contributions, in eqn(\ref{final1}). 



\subsection{Noncommutative case $(g=\ep\eta)$}\label{cse2}

We now turn to the Seiberg Witten limit, (\ref{swl}) which gives rise to 
noncommutative field theory on the brane. Here again we will be 
interested in writing out the two point function eqn(\ref{clchannel}) in 
the closed string 
channel as a sum of the massless closed string modes. Due to the scaling 
of the closed string metric, unlike the earlier case, we 
will now expand all results in powers of the scale 
parameter for closed string metric, $\ep$. We begin by expanding the DBI 
action,

\beqa
{\cal 
L}=\sqrt{(2\pi\al)e^{-(P+Q)\phi}(B+b)}\left[1+\f{1}{(2\pi\al)e^{-Q\phi}(B+b)}
\ep(\eta+h)\right]^{1/2}
\eeqa

The $\phi$ and $b$-field vertices from this are,

\beqa
V_{\phi}=\sqrt{(2\pi\al)B}\left[-\f{1}{2}\left(\f{1}{B}\right)^{kl}
+\f{\ep^2(4Q-2)}{4(2\pi\al)^2}\left(\left(\f{1}{B^3}\right)^{kl}
-\f{1}{4}\Tr\left(\f{1}{B^2}\right)\left(\f{1}{B}\right)^{kl}\right)\right]
\nonumber
\eeqa
\beqa
V_b=\sqrt{(2\pi\al)B}\left[\f{1}{4}\left(\f{1}{B}\right)^{kl}
\left(\f{1}{B}\right)^{ji}-\f{1}{2}\left(\f{1}{B}\right)^{jl}
\left(\f{1}{B}\right)^{ki}
\right]
\eeqa

Note that We are interested in the two point function only upto 
${\cal O}(\ep^2)$, hence we need not consider the graviton vertex. 
Also the $b$-field propagator has a $\ep^2$ factor (\ref{bp}). So, it 
is only necessary to compute the dilaton vertex upto ${\cal O}(\ep^2)$.

\begin{figure}[t]
\begin{center}
\begin{psfrags}
\psfrag{B1}[][]{\footnotesize{$1/B$} }
\psfrag{B2}[][]{\footnotesize{$1/B^2$} }
\psfrag{B3}[][]{\footnotesize{$1/B^3$} }
\psfrag{F}[][]{\footnotesize{$F$} }
\psfrag{b}[][]{$b$ }
\psfrag{p}[][]{$\phi$ }
\psfrag{i}[][]{\footnotesize{(i)}}
\psfrag{ii}[][]{\footnotesize{(ii)}}
\psfrag{iii}[][]{\footnotesize{(iii)}}
\epsfig{file=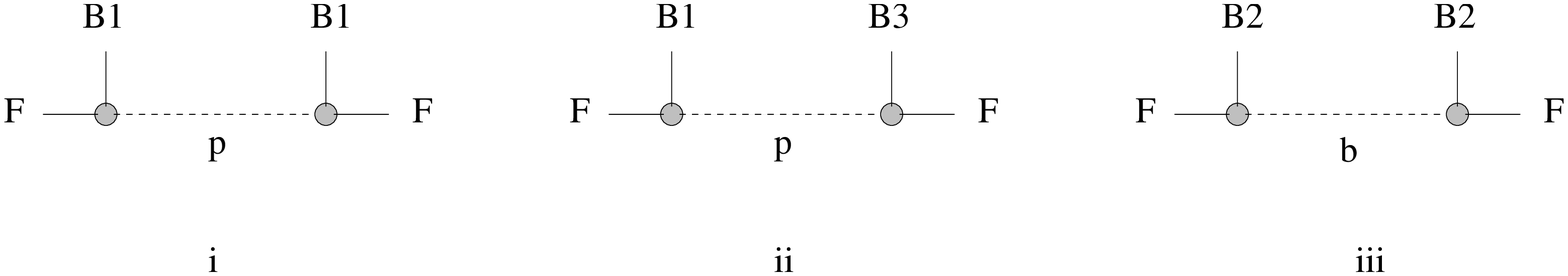, width= 12 cm,angle=0}
\end{psfrags}
\vspace{ .1 in }
\caption
{Two point amplitude upto ${\cal O}(\ep^2)$. (i) and (ii) are due to
dilaton exchange, (iii) is due to $b$-field exchange.}
\label{gepe}
\end{center}
\end{figure}

The exchanges with these couplings are summarised in Figure~\ref{gepe}.
The propagators in this limit, eqns(\ref{dp2},\ref{bp}),

\beqa
<\phi\phi>&=&-\f{(D-2)i\ka^2}{4}\f{1}{\kpe^2+\ep^{-1}\kpa^2}\\
<b_{ij}b_{i^{'}j^{'}}>&=&-\f{2i\ka^2\ep^2}{(2\pi\al)^2}
\f{[\eta_{ii^{'}}\eta_{jj^{'}}-\eta_{ji^{'}}
\eta_{ij^{'}}]}{\kpe^2+\ep^{-1}\kpa^2}
\eeqa

With the vertices computed above and the propagator in this limit, 
the two point function is,

\beqal{final2}
L_2&=&-i\mbox{det}(2\pi\al B)\ka^2\tau_p^2
\int\f{d^l\kpe}{(2\pi)^l}\f{1}{\kpe^2+\ep^{-1}p^2}
[{\cal O}(1)+{\cal O}(\ep^2)]
\eeqa

\noindent
where,

\beqa
{\cal O}(1)=
\left[\f{(D-2)}{32}
\left(\f{1}{B}\right)^{kl}\left(\f{1}{B}\right)^{k^{'}l^{'}}
+(kl) \leftrightarrow (k^{'}l^{'})\right]
\eeqa

\beqa
{\cal O}(\ep^2)&=& 
\f{\ep^2}{(2\pi\al)^2}\f{(D-2)}{16}\left[\left[\left(\f{1}{B^3}\right)^{kl}
-\f{1}{4}\Tr\left(\f{1}{B^2}\right)\left(\f{1}{B}\right)^{kl}\right]
\left(\f{1}{B}\right)^{k^{'}l^{'}}\right]\non 
&+&\f{\ep^2}{(2\pi\al)^2}\left[
\f{1}{4}\left(\f{1}{B^2}\right)^{kk^{'}}\left(\f{1}{B^2}\right)^{ll^{'}}
-\f{1}{4}\left(\f{1}{B^2}\right)^{k^{'}l}\left(\f{1}{B^2}\right)^{kl^{'}}
\right]\non
&+& (kl) \leftrightarrow (k^{'}l^{'})
\eeqa

\noindent
We can now reconstruct the quadratic term in effective action, 
(\ref{eff}) following the earlier case.
With the following expansions, it is easy to check that the sum of the 
massless contributions adds upto eqn(\ref{clchannel}).

\beqal{expan2}
G^{ij}&\sim &-\f{\ep}{(2\pi\al)^2}\left(\f{1}{B^2}\right)^{ij} +{\cal
O}(\ep^3)\\
\th^{ij}&\sim &
\left(\f{1}{B}\right)^{ij}+ 
\f{\ep^2}{(2\pi\al)^2}\left(\f{1}{B^3}\right)^{ij}\non
\sqrt{\ep\eta+(2\pi\al)B}&\sim &
\sqrt{(2\pi\al)B}\left[1-\f{\ep^2}{4(2\pi\al)^2}\Tr\left(\f{1}{B^2}\right)
\right]\nonumber
\eeqa

\noindent
Note that, at the tree-level, to the linear order, $\hat{F}=F$, 
(\ref{redeff}). At this 
quadratic order in the effective action there is no need for 
redefinition of $F$ to equate the result here with that of string theory 
result in eqn(\ref{clchannel}).

\subsection{Noncommutative case ($G=\eta$)}\label{cse3}

In this part we finally consider the restriction of the open string 
metric, $G=\eta$.
The lowest order solution for the closed string metric, $g$ in $\al$ in 
this limit is,

\beqal{soln}
g=-(2\pi\al)^2B^2 +{\cal O}(\alpha^{'4})
\eeqa

We will now consider expansions of the two point functions in 
powers of $\al$.
We begin again with the following DBI Lagrangian,

\beqa
{\cal L}&=&\sqrt{(2\pi\al)e^{-(P+Q)\phi}(B+b)}
\left[1-\f{1}{e^{-Q\phi}(B+b)}(2\pi\al)B^2(\eta+h)^2\right]^{1/2}\non
\eeqa

The calculation for the vertices is same as before, there is no graviton 
vertex to the leading orders. The dilaton and the $b$-field vertices 
are,

\beqa
V_{\phi}=\sqrt{(2\pi\al)B}\left[-\f{1}{2}\left(\f{1}{B}\right)^{kl}+
\f{(2\pi\al)^2(4Q-2)}{4}\left(B^{kl}-\f{1}{4}\Tr(B^2)\left(\f{1}{B}\right)^{kl}
\right)\right]\nonumber
\eeqa
\beqa
V_b=\sqrt{(2\pi\al)B}\left[\f{1}{4}\left(\f{1}{B}\right)^{kl}
\left(\f{1}{B}\right)^{ji}-\f{1}{2}\left(\f{1}{B}\right)^{jl}
\left(\f{1}{B}\right)^{ki}
\right]
\eeqa

The propagators for the dilaton and the $b$-field are modified as,

\beqa
<\phi\phi>&=&-\f{(D-2)i\ka^2}{4}\f{1}{\kpe^2+\tilde{\kpa}^2/(2\pi\al)^2}\\
<b_{ij}b_{i^{'}j^{'}}>&=&-2i\ka^2(2\pi\al)^2
\f{[B^2_{ii^{'}}B^2_{jj^{'}}-B^2_{ji^{'}}
B^2_{ij^{'}}]}{\kpe^2+\tilde{\kpa}^2/(2\pi\al)^2}
\eeqa

\begin{figure}[t]
\begin{center}
\begin{psfrags}
\psfrag{B1}[][]{\footnotesize{$1/B$} }
\psfrag{B2}[][]{\footnotesize{$1/B^2$} }
\psfrag{B}[][]{\footnotesize{$B$} }
\psfrag{F}[][]{\footnotesize{$F$} }
\psfrag{b}[][]{$b$ }
\psfrag{p}[][]{$\phi$ }
\psfrag{i}[][]{\footnotesize{(i)}}
\psfrag{ii}[][]{\footnotesize{(ii)}}
\psfrag{iii}[][]{\footnotesize{(iii)}}
\epsfig{file=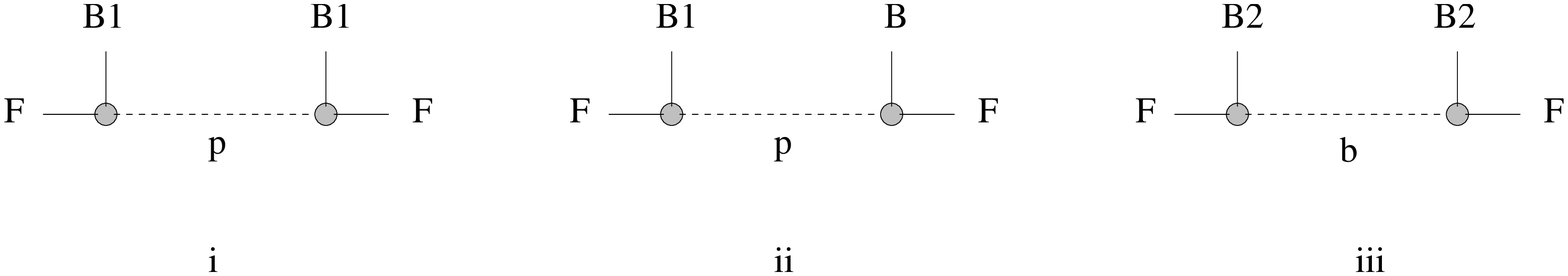, width= 12 cm,angle=0}
\end{psfrags}
\vspace{ .1 in }
\caption
{Two point amplitude upto ${\cal O}(\als)$. (i) and (ii) are due to
dilaton exchange, (iii) is due to $b$-field exchange.}
\label{ge}
\end{center}
\end{figure}

With these vertices (shown in Figure~\ref{ge}) and the propagators from 
eqns(\ref{dp3},\ref{bp}), 
the two point function is now given by,

\beqal{final3}
L_2=-i\det(2\pi\al 
B)\ka^2\tau_p^2\int 
\f{d^l\kpe}{(2\pi)^l}\f{1}{\kpe^2+\tilde{p}^2(2\pi\al)^2}[{\cal O}(1)
+{\cal O}(\als)]
\eeqa

\beqa
{\cal O}(1)=
\left[\f{(D-2)}{32}
\left(\f{1}{B}\right)^{kl}\left(\f{1}{B}\right)^{k^{'}l^{'}}
+(kl) \leftrightarrow (k^{'}l^{'})\right] 
\eeqa

\beqa
{\cal O}(\als)&=&
(2\pi\al)^2\f{(D-2)}{16}\left[\left[B^{kl}
-\f{1}{4}\Tr(B^2)\left(\f{1}{B}\right)^{kl}\right]
\left(\f{1}{B}\right)^{k^{'}l^{'}}\right] \non
&+&(2\pi\al)^2\left[
\f{1}{4}\left(\eta^{ll^{'}}\eta^{kk^{'}}
-\eta^{kl^{'}}\eta^{lk^{'}}\right)
\right]\non
&+& (kl) \leftrightarrow (k^{'}l^{'})
\eeqa

\noindent
We will need the following expansions in this limit, to 
expand the closed string channel result upto this order.
We have already set,

\beqa
G^{ij}&=&\eta^{ij}
\eeqa

\noindent
and with the solution for $g$, eqn(\ref{soln}) to the lowest order in 
$\al$,

\beqal{expan3}
\th^{ij}&\sim & \left(\f{1}{B}\right)^{ij}+(2\pi\al)^2B^{ij}\\
\sqrt{g+(2\pi\al)B}&\sim&
\sqrt{(2\pi\al)B}\left[1-\f{(2\pi\al)^2}{4}\Tr(B^2)\right]
\eeqa

\noindent
As in the earlier cases, the massless contributions computed here, 
eqn(\ref{final3}) adds upto eqn(\ref{clchannel}).
Note that the situation here is similar to that of the 
earlier case in Section~\ref{cse2}. As $\al \sim \sqrt{\ep}$, the 
closed string metric in both the cases goes to zero as $g\sim \ep$.
However the difference being that the two point amplitude differ
by powers of $B$ in both the cases, due to the  
relative power of $B^2$ in $g$ in this case. Here too, the SW 
map 
between the usual and the noncommutative field strength 
eqn(\ref{redeff}), remains the same. The differences in the powers of 
$B$ in the two point amplitudes, eqn(\ref{final2}) and 
eqn(\ref{final3}) are absorbed in $G$, $\theta$ and 
$\sqrt{g+(2\pi\al)B}$ in the two cases. We can work with any of the 
forms of the closed string metric $g$, the important point being that 
$g$ should go to zero as $\ep$ which gives the noncommutative gauge 
theory on the brane.

\section{Open superstring in background $B$-field}\label{osbf}

\begin{figure}[t]
\begin{center}
\begin{psfrags}
\psfrag{orb}[][]{$C^2/Z_2$}
\psfrag{d3}[][]{$D_3$ Brane}
\psfrag{b}[][]{$B$ Field}
\epsfig{file=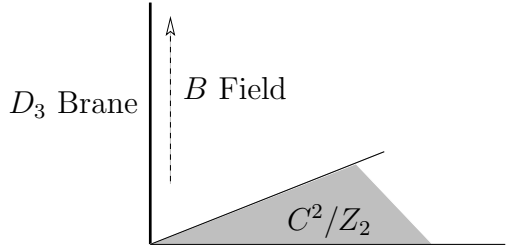, width= 6cm,angle=0}
\end{psfrags}
\vspace{ .1 in }
\caption{The background for string propagation. $D_3$ brane world volume 
directions are 0,1,2,3. Orbifolded directions 6,7,8,9.}
\label{back}
\end{center}
\end{figure}

After studying the bosonic case, let us now turn to the supersymmetric 
example. With the observations made in Section~\ref{osol} we 
will consider ${\cal N}=2$ gauge theory that can be realised on a 
{\it fractional} $D_3$ brane located at the fixed point of $C^2/Z_2$ 
orbifold. The setup is shown in Figure~\ref{back}. With the $B$-field 
turned on 
along the world-volume directions, the low-energy effective theory on 
the brane is described by a noncommutative ${\cal N}=2$ gauge theory.
The world-sheet action for the fermions coupled to $B$-field is given 
by,

\beqal{sf}
S_F=\f{i}{4\pi\al}\int_{\Sigma}g_{MN}\bps^{M}\rho^{\alpha}\pa_{\alpha}
\ps^{N}-\f{i}{4}\int_{\pa\Sigma}B_{MN}\bps^{N}\rho^0\ps^{M}
\eeqa

The full action including the bosons (\ref{sb}) and the fermions 
(\ref{sf}) with the bulk and 
the boundary terms are invariant under the following supersymmetry 
transformations,

\beqa
\de X^M&=&\bar{\ep}\psi^M\non
\de \psi^M&=&-i\rho^{\alp}\pa_{\alp}X^M\ep
\eeqa

We now write down the boundary equations by varying (\ref{sf})
with the following constraints,

\beqa
\de\ps^{M}_L=\de\ps^{M}_R\mid_{\s=\pi}
\mbox{\hspace{0.2in}and\hspace{0.2in}}
\de\ps^{M}_L=-(-1)^a\de\ps^{M}_R\mid_{\s=0}
\eeqa

where, $a=0,1$ gives the NS and the R sectors respectively
This gives the following boundary equations,

\beqal{bcf}
g_{MN}(\ps_L^N-\ps_R^N)+2\pi\al B_{MN}(\ps_L^N+\ps_R^N)=0
\mid_{\s=\pi}\\
g_{MN}(\ps_L^M+(-1)^a\ps_R^M) + 2\pi\al B_{MN}(\ps_L^N-(-1)^a\ps_R^N)=0
\mid_{\s=0}
\eeqa

\noindent
To write down the correlator for the fermions, first
define,

\beqa
\psi^M&=&\psi^M_L(\s,\tau) \mbox{\hspace{0.5in}} 0\leq\s\leq\pi\non 
&=&\left(\f{g-2\pi\al B}{g+2\pi\al B}\right)^M_N\psi_R^N
(2\pi-\s,\tau) \mbox{\hspace{0.5in}} \pi\leq\s \leq 2\pi
\eeqa
\noindent
This is the usual doubling trick that ensures the boundary conditions 
(\ref{bcf}). 
In the following 
section we would compute the two point function for the gauge field on 
the brane by inserting two vertex operators at the boundaries of the 
cylinder. Restricting ourselves to the directions along the brane, this 
vertex operator for the gauge field in the zero picture is given by,

\beqa
V(p,x,y)=\f{g_o}{(2\al)^{1/2}}\ep_j\left(i\pa_y X^j+4p.\Psi\Psi^j\right)
e^{ip.X}(x,y)
\eeqa

where $\Psi^i$ is given by,
\beqal{Psi}
\Psi^i(0,\tau)&=&\f{1}{2}\left(\psi^i_L(0,\tau)+(-1)^{a+1}\psi^i_R(0,\tau)\right)
=\left(\f{1}{g-2\pi\al B}g\right)^{i}_{j}\psi_L^j(0,\tau)\non
\Psi^i(\pi,\tau)&=&\f{1}{2}\left(\psi^i_L(\pi,\tau)+ 
\psi^i_R(\pi,\tau)\right)
=\left(\f{1}{g-2\pi\al B}g\right)^{i}_{j}\psi_L^j(\pi,\tau)
\eeqa

\noindent
Using (\ref{Psi}), the correlation function for $\Psi$ is given by,

\beqa
\expect{\Psi^i(w)\Psi^j(w^{'})}=G^{ij}{\cal G}\alpb (w-w^{'})
\eeqa

\noindent
where $G^{ij}$ is the open string metric defined in (\ref{gth}) and,
${\cal G}\alpb (w-w^{'})$ is given by \cite{narain},

\beqa
{\cal G}\alpb (w-w^{'})
=\f{\al}{4\pi}\f{\vth\alpb\left(\f{w-w^{'}}{2\pi},it\right)
\vth^{'}\oo (0,it)}{\vth\oo\left(\f{w-w^{'}}{2\pi},it\right)
\vth\alpb (0,it)}
\eeqa

\noindent
$\alp,\beta$ denotes the spin structures. $\alp=(0,1/2)$ are the NS 
and the R sectors and $\beta=(0,1/2)$ stands for the 
absence or the presence of the world-sheet fermion number $(-1)^F$ with 
$\psi$ being anti-periodic or periodic along the $\tau$ direction on the 
world-sheet. $w$, $w^{'}$ are located at the boundaries of the 
world-sheet for the open string which is a cylinder, i.e. at $\s=0,\pi$.

\subsection{Strings on $C^2/Z_2$ orbifold}

An efficient and simple way to break ${\cal N}=4$ supersymmetry and  
obtain gauge theories with less supersymmetries is by orbifolding the 
background space. Specifically strings on $C^2/Z_2$ gives rise to ${\cal 
N}=2$ supersymmetric gauge theory on D3-branes with world volume 
directions transverse to the Orbifolded planes. The open and the closed 
string spectrum on this orbifold have been nicely worked out in 
\cite{Bertolini:2001gq}. We 
include a brief analysis here that will be relevant in the later 
discussions.  
We will take the orbifolded directions to be $6,7,8,9$ with 
$Z_2=\{g_i\mid e,g\}$ such that $g^2=e$
The action of $g$ on these coordinates is given by,

\beqa
gX^I=-X^I \mbox{\hspace{0.5in} for \hspace{0.5in}} I=6,7,8,9
\eeqa

\noindent
In order to preserve world sheet supersymmetry we must also consider the 
action of $Z_2$ on the fermionic partners, $\psi^I$.

\subsubsection{Open string spectrum and fractional branes}\label{oss}

On a particular
state of the open string the orbifold action is on the oscillators, 
$\psi^I_{-r}$ along
with the Chan-Paton indices associated with it. Let us consider the
massless bosonic states from the NS sector.

\beqa
g|i,j,\psi^I_{-1/2}>=\gamma_{ii^{'}}|i^{'},j^{'},\hat{g}\psi^I_{-1/2}>
\gamma^{-1}_{j^{'}j}
\eeqa

where $\gamma$ is a representation of $Z_2$
The spectrum is obtained by keeping the states that are invariant under 
the above action. To derive this it is easier to work in the basis 
where $\gamma$ is diagonal,

\beqa
\gamma=\s_3=\left(\begin{matrix} 1&0\\0&-1 \end{matrix}\right)
\eeqa

The action on the Chan-Paton indices can be thought of as,

\beqa
\gamma\left(\begin{matrix} 11&12\\21&22 \end{matrix}\right)\gamma^{-1}
=\left(\begin{matrix} 11&-12\\-21&22 \end{matrix}\right)
\eeqa

Thus the diagonal ones survive for the $Z_2$ action on the oscillators 
is $\hat{g}|\psi^I_{-1/2}>=|\psi^I_{-1/2}>$, i,e. for $I=2,3,4,5$ and 
the 
off-diagonal ones are preserved for oscillators that are odd under the 
$Z_2$ action, $\hat{g}|\psi^I_{-1/2}>=-|\psi^I_{-1/2}>$, i,e. for 
$I=6,7,8,9$. The spectrum can thus be summarised as,

\beqa
A^I\rightarrow 
\left[\begin{matrix} A^{I}_1&A^{I}_2& \mbox{$2$ gauge fields}&I=2,3\\
                \phi^{I}_1&\phi^{I}_2&\mbox{$4$ real scalars}&I=4,5\\
             \Phi^{I}_1&\Phi^{I}_2&\mbox{$8$ real scalars}&I=6,7,8,9\\
\end{matrix}\right]
\eeqa

\noindent
The orbifold action on the space-time spinors is given by,

\beqal{spin}
\chi_{ij} \rightarrow \gamma_{ii^{'}}e^{i\pi(s_3+s_4)}\chi_{i^{'}j^{'}}
\gamma_{j^{'}j}^{-1}
\eeqa

\noindent
where $\chi_{i^{'}j^{'}}=\ket{s_1,s_2,s_3,s_4}_{i^{'}j^{'}}$ are the 
sixteen spinors 
of 
$SO(8)$, with $s_i=\pm 1/2$. The spinors in (\ref{spin}) are left 
invariant for, $s_3+s_4=0$ and $s_3+s_4=\pm 1$. The first one leaves 
$\chi_{11}$ and $\chi_{22}$ invariant and the second one leaves 
$\chi_{12}$ and $\chi_{21}$. Projection onto one of the chiralities 
leaves four copies of each of the spinors.

The above fields can be grouped into two vector multiplets and two 
hypermultiplets of ${\cal N}=2
$ with gauge group $U(1)\times 
U(1)$.
The beta function for the gauge couplings for this theory vanishes and 
the theory is conformally invariant. Now consider an irreducible 
representation $\gamma=\pm 1$. This acts 
trivially on the Chan-Paton indices. These branes are known as {\it 
fractional branes}. From the geometric point of view there is no image 
for the $D3$
brane. The brane is localised at the fixed point on the orbifold 
plane but is free to move in the non-orbifolded (4,5) directions. 

Following the above analysis, the spectrum consists of 
a single gauge field and two scalars completing the vector multiplet
of ${\cal N}=2$ with gauge group  $U(1)$. The beta function for this 
theory is nonzero.  With a constant 
$B$-field turned on along the world volume directions of the 
$D_3$-brane,$(0,1,2,3)$, the low energy dynamics on the brane will be 
described be noncommutative gauge theory in the Seiberg-Witten limit. 
In the following section we will study the
ultraviolet behaviour of this theory and see how the UV divergences have 
a natural interpretation in terms of IR divergences due to massless 
closed string modes as a result of open-closed string duality.

\subsubsection{Closed string spectrum}\label{css}

The closed string theory consists of additional twisted sectors apart 
from the untwisted sectors. The orbifold action on the space-time 
implies the following boundary conditions on the world-sheet bosons and 
fermions,

\beqa
X^{I}(\s+2\pi,\tau)&=&\pm X^{I}(\s,\tau)\non
\psi^{I}(\s+2\pi,\tau)&=&\pm \psi^{I}(\s,\tau)
\mbox{\hspace{.2in}} I=6,7,8,9
\eeqa

For the world sheet fermions, the $(+)$-sign stands for the NS-sector 
and the $(-)$-sign for the R-sector. For the other directions the boundary 
conditions on the world-sheet fields are as usual. We will first list 
the fields in the untwisted sectors. In the NS-NS sector the massless 
states invariant under the orbifold projection are,

\beqa
\psi_{-1/2}^{I}\tilde{\psi}_{-1/2}^{J}\ket{0,k}  
\eeqa

where, $I,J=\{2,3,4,5\}$ or $I,J=\{6,7,8,9$\}. The first set of 
oscillators give the graviton, antisymmetric 2-form field, and the 
dilaton. The second set gives sixteen scalars.

The orbifold action on the spinor of $SO(8)$ is given by,

\beqa
\ket{s_1,s_2,s_3,s_4} \rightarrow 
e^{i\pi(s_3+s_4)}\ket{s_1,s_2,s_3,s_4}
\eeqa

The $Z_2$ invariant R-R state is formed by taking both the left 
the right states to be either even or odd under $Z_2$ projection
corresponding to $s_3+s_4=0$ or $s_3+s_4=\pm 1$ respectively. GSO 
projection, restricting to both the left and right states to be of the 
same chirality gives thirty two states. These states correspond to four 
2-form fields and eight scalars.

Let us now turn to the twisted sectors. 
For the twisted sectors the ground state energy for both the
NS and the R sectors vanish. In the NS sector the massless modes
come from $\psi^I_0$, $I=6,7,8,9$ oscillators which form a spinor
representation of $SO(4)$. With the GSO and the orbifold projections,
the closed string spectrum is given by, $2\times 2 =[0]+[2]$. The [0]
and the self-dual [2] constitute the four massless scalars in the
NS-NS
sector. Similarly, in the R sector, the massless modes are given by
$\psi^I_0$ for
$I=2,3,4,5$. Thus giving a scalar and a two-form self-dual field in the
closed string R-R sector. The couplings for the massless closed string
states to the fractional $D_3$-brane will be studied in Sections~\ref{ons} and 
\ref{orr}.

\section{Open string one loop amplitude : IIB on $C^2/Z_2$}\label{tpa}

In this section we compute the two point function for the gauge fields 
on the brane. The necessary ingredients are given in Section~\ref{osbf} and in 
the Appendix~\ref{vacuum}. The one loop vacuum amplitude  
vanishes as a result of supersymmetry, i.e.

\beqal{zpt}
\det(g+2\pi\al B)\int^{\infty}_{0}\f{dt}{4t}(8\pi^2\al t)^{-2}
\sum_{(\alp,\beta,g_i)}Z\alpb_{g_i}=0 
\eeqa

\noindent
The factor of $\det(g+2\pi\al B)$ comes from the trace over the 
world sheet bosonic zero modes.
The sum is over the spin structures $(\alp,\beta)=(0,1/2)$ corresponding 
to the NS and R sectors 
and the GSO projection and the orbifold projection. The elements 
$Z_{g_i} {\alpb}$ are computed in the 
Appendix~\ref{vacuum}. Let us now compute the two point function. This 
is given by,

\beqal{2pt}
A(p,-p)&=&
\det(g+2\pi\al B)\int^{\infty}_{0}\f{dt}{4t}(8\pi^2\al t)^{-2}
\times \non &\times&\sum_{(\alp,\beta,g_i)}Z {\alpb}_{g_i}
\int_{0}^{2\pi t}dy\int_{0}^{2\pi t}dy^{'}
\expect{V(p,x,y)V(-p,x^{'},y^{'})}_{(\alp,\beta)}\non
\eeqa

\noindent
For the flat space, it is well known that amplitudes with less that four 
boson insertions vanish. However, in this model the two point amplitude 
survives. We will now compute this amplitude in the presence of 
background $B$-field. First note that the bosonic correlation function, 
$\expect{:\pa_yX^ie^{ip.X}::\pa_y^{'}X^ie^{-ip.X}:}$, does not 
contribute to 
the two point amplitude as it is independent of the spin structure. The 
two point function would involve the sum over the $Z_{g_i} \alpb$ which 
makes this contribution zero due to (\ref{zpt}). The nonzero 
part of the amplitude will be obtained 
from the fermionic part,

\beqa
\ep_k\ep_l\expect{:p.\Psi\Psi^ke^{ip.X}::
p.\Psi\Psi^le^{-ip.X}:}&=&\ep_k\ep_l p_ip_j
\left(G^{il}G^{jk}-G^{ij}G^{kl}\right)\times\\
&\times&{\cal G}^2\alpb (w-w^{'}) \expect{:e^{ip.X}::e^{-ip.X}:}\nn
\eeqa

\noindent
For the planar two point amplitude, both the vertex operators would be 
inserted at the same end of the cylinder (i.e. at $w=0+iy$ or $\pi+iy$). 
In this case, the sum in the two point amplitude reduces to,

\beqal{ssum}
\sum_{(\alp,\beta,g_i)}Z\alpb_{g_i}{\cal G}^2\alpb(i\De y/2\pi)&=&
\sum_{(\alp,\beta)}Z\alpb_e{\cal G}^2\alpb(i\De y/2\pi)\non &+&
\sum_{(\alp,\beta)}Z\alpb_g{\cal G}^2\alpb(i\De y/2\pi)\nn
\eeqa
\beqa
&=&\f{4\pi^2}{\eta(it)^{6}\vth^2_1(i\De y/2\pi,it)}\sum_{(\alp,\beta)}
\vth^2\alpb(0,it)\vth^2\alpb(i\De y/2\pi,it)+ \\
&+& \f{16\pi^2}{\vth^{2}_{1}(i\De y/2\pi,it)
\vth^2_2(0,it)}\left[\vth^2_3(i\De
y/2\pi,it)\vth^{2}_{4}(0,it)-
\vth^2_4(i\De y/2\pi,it)\vth^{2}_{3}(0,it)\right]\nn
\eeqa

\noindent
where, $\De y=y-y^{'}$.
We have separated the total sum as the sum over the two $Z_2$ 
group 
actions. In writing this we have used the following identity 

\beqa\eta(it)=\left[\f{\pa_{\nu}\vth_1(\nu,it)}{-2\pi}\right]^{1/3}_{\nu=0}
\eeqa

\noindent
Now, the first term vanishes due to the following identity

\beqa
\lefteqn{\sum_{(\alp,\beta)}\vth\alpb (u)\vth\alpb (v)\vth\alpb (w)\vth\alpb (s)
= }\non & & 2\vth\oo (u_1)\vth\oo (v_1)\vth\oo (w_1)\vth\oo (s_1)\non
\eeqa

\noindent
where,
\beqa
u_1&=&\f{1}{2}(u+v+w+s) \mbox{\hspace{0.2in}} v_1=\f{1}{2}(u+v-w-s) 
\mbox{\hspace{0.2in}}\non w_1&=&\f{1}{2}(u-v+w-s) \mbox{\hspace{0.2in}}
s_1=\f{1}{2}(u-v-w+s)
\eeqa

\noindent
and noting that, $\vth\oo(0,it)=0$,
in the same way as the flat case that makes amplitudes with two 
vertex insertions vanish. The second term is a constant also due to,

\beqal{ident}
\vth^2_4(z,it)\vth^2_3(0,it)-\vth^2_3(z,it)\vth^2_4(0,it)
=\vth^2_1(z,it)\vth^2_2(0,it)
\eeqa

\noindent
For the nonplanar amplitude, that we are ultimately interested in, we 
need to put the two vertices at the two ends of the cylinder such that, 
$w=\pi+iy$ and $w^{'}=iy^{'}$. It can be seen that the fermionic part
of the correlator is constant and independent of $t$, same as the 
planar case following from the identity (\ref{ident}). The effect of 
nonplanarity 
and the regulation of the two point function due to the background 
$B$-field is encoded in the correlation functions for the exponentials. 
The two point function thus reduces to,

\beqa
A(p,-p)\sim \ep_k\ep_l p_ip_j
\left(G^{il}G^{jk}-G^{ij}G^{kl}\right)\int^{\infty}_{0}\f{dt}{4t}(8\pi^2\al 
t)^{-2}\int_{0}^{2\pi
t}dy dy^{'}\expect{e^{ip.X}e^{-ip.X}}\non
\eeqa

\noindent
The noncommutative gauge theory two point function is obtained in the 
limit
$t \rightarrow \infty$ and $\al \rightarrow 0$. The correlation function 
in this limit can be 
computed from the bosonic correlation functions 
\cite{oneloop3,oneloop4,oneloop5,Chaudhuri,callan}.
For the nonplanar case in this limit, 

\beqal{ti}
\expect{e^{ip.X}e^{-ip.X}}&=& \exp\left\{-p^2t \De x(\De x-1)-\f{1}{4t}
p_i(g^{-1}-G^{-1})^{ij}p_j\right\}\non &=&\exp\left\{-p^2t \De x(\De 
x-1)-\f{\tilde{p}^2}{4t}\right\}
\eeqa 

\noindent
where, $\tilde{p}=(\th p)$. We have redefined the world sheet coordinate 
as $\De x =\De y/(2\pi 
t)$ and have scaled $t \rightarrow t/(2\pi\al)$. We have also used the 
following relation in writing down the last expression.

\beqa
g^{-1}=G^{-1}-\f{(\th G \th)}{(2\pi\al)^2}
\eeqa

\noindent
The first term in the exponential in (\ref{ti}) regulates the integral 
over $t$ in the infrared, for $p\neq 0$ and the 
second term regulates it in the ultraviolet that is 
usually observed in noncommutative field theories.
The $t \rightarrow \infty$ limit suppresses the contributions from all 
the open 
string massive modes. However as, discussed in Section~\ref{osol},  
the field theory divergences still come from the $t\rightarrow 0$ 
region. We can 
thus break the integral over $t$ into two intervals 
$1/\Lambda^2\al<t<\infty $ and $0<t<1/\Lambda^2\al$ (see 
Figure~\ref{lim1}). The 
second interval 
which is the source of the UV divergence is also the regime dominated by 
massless closed string exchanges. We now evaluate the two 
point function in this limit. First, the correlation function for the 
exponential in the $t \rightarrow 0$ limit is given by

\beqal{tz}
\expect{e^{ip.X}e^{-ip.X}}=\exp\left\{-\f{\al\pi}{2t}p_ig^{ij}p_j\right\}
\eeqa

\noindent
where $g^{ij}$ is the closed string metric. 
Modular transformation , ($t\rightarrow 1/t$) allows us 
to rewrite the 
one loop amplitude as the sum over closed string modes in a tree 
diagram.
In the limit $t\rightarrow 0$, the amplitude will be 
dominated by massless closed string modes.
In this model however, the effect of the massive modes in the loop 
cancel amongst themselves for any value of $t$. In the open string 
channel the $t \rightarrow 0$ limit would usually be contributed by the 
full tower of 
open string modes. However since we have seen that the effect of the 
massive string modes cancel anyhow for all values of $t$, the 
contribution to this limit from the open string modes comes only from 
the massless ones. The additional term in (\ref{ti}) as compared to 
(\ref{tz}) gives finite derivative corrections to the effective action. 
These would in general require the massive closed string states for its 
dual description. Without these derivative corrections, the 
contributions from the massless open string loop and the massless closed 
string tree are exactly equal.
The divergent ultraviolet 
behaviour of the massless open string modes can thus be captured by the 
the massless closed string modes that have momentum in the limit 
$[0,1/\Lambda\al]$. The amplitude can now be written as,

\beqal{sum}
A(p,-p)={\cal V}_4\det(g+2\pi\al B)\left(\f{g_o^2}{8\pi^2\al}\right)
\ep_k\ep_l p_ip_j
\left(G^{il}G^{jk}-G^{ij}G^{kl}\right)I(p)\non
\eeqa

\noindent
where,

\beqal{ip}
I(p)&=&\int ds s^{-1} 
\exp\left\{-\f{\al\pi s}{2}p_ig^{ij}p_j\right\}\non
&=&4\pi\int \f{d^2\kpe}{(2\pi)^2}\f{1}{\kpe^2+p_ig^{ij}p_j}
\eeqa

\noindent
The integral is written in terms of $s=1/t$ and in the last line we have 
rewritten it as an integral over $\kpe$, the momentum in the 
directions transverse to the brane for 
closed strings. The nonzero contribution to the two point amplitude 
in (\ref{2pt}) comes from the $\Tr_{NS}\left[gq^{L_0}\right]$ and 
$\Tr_{NS}\left[g(-1)^F q^{L_0}\right]$, that are evaluated
in (\ref{part}). These correspond to anti-periodic NS-NS and periodic 
(R-R) 
closed strings in the twisted sectors respectively. The fractional 
$D_3$-brane is localised at the fixed point of $C^2/Z_2$. Thus the 
twisted sector closed string states that couple to it are twisted in 
all the directions of the orbifold. 
These modes are localised at the fixed point and are free to move in the 
six directions transverse to the orbifold. This is the origin of the 
momentum integral (\ref{ip}) in two directions transverse to the 
$D$-brane. These twisted states come from both the NS-NS and the R-R 
sectors and are listed in Section~\ref{css}.

As we are interested in seeing the ultraviolet effect of the 
open 
string channel as an infrared effect in the closed string channel, 
like in the bosonic case (see eqn.(\ref{cut-off})), we 
must cut off the $s$ integral at the lower end at some value 
$\Lambda^2\al$ corresponding to the UV cut-off for the momentum 
of the massless closed strings in the directions transverse to the brane.
With this, we have,



\beqal{final}
I(p,\Lambda)=4\pi^2\ln\left(\f{p_ig^{ij}p_j+1/(\Lambda\al)^2}
{p_ig^{ij}p_j}\right)
\eeqa

\noindent
This is the ultraviolet behaviour of the two point function for two gauge 
fields in 
${\cal N}=2$ theory. For the noncommutative theory it is regulated for 
$p\neq 0$. The fact that we are able to rewrite the gauge theory two 
point function as massless closed string tree-level exchanges is very 
specific to the ${\cal N}=2$ theory. The computations above show that 
the origin of this can be traced to open-closed string duality where the 
orbifold background cancels all contributions from the 
massive states as far as the UV singular terms are concerned. The 
background $B$-field in the SW limit only acts as a physical 
regulator.

\section{Massless closed string exchanges}\label{ocse}

Let us now study in detail the massless closed string exchanges as in 
Section~\ref{cse} that is discussed in more detail in \cite{myopcl3}. 
The computation will be done for the three cases enumerated at the begining of
this section. We will calculate the contribution to the nonplanar two point 
function with two gauge fields on the brane with massless closed string 
exchanges coming from the NS-NS and the R-R sectors. First we study the flat 
10D case and then we will move on to the exchanges in the $C^2/Z_2$ orbifold 
background.

\subsection{Type IIB on flat space}\label{flat}

In this section we will consider the flat 10D case and consider massless 
closed string exchanges for a $D_3$ brane. The two point function will be 
shown to vanish as expected from the analysis in the previous section. See 
eqn(\ref{ssum}). The results here will be necessary for the later part of 
this section when we study the exchanges on $C^2/Z_2$ orbifold. These will 
precisely be the contributions from the untwisted states upto an overall 
constant.
To begin we write down the supergravity action for the type IIB theory
in the Einstein frame \footnote{$(A_p\w B_q)_{i_1...i_p j_1...j_q}=\f{(p+q)!}{p!q!}A_{[i_1...i_p}B_{j_1...j_q]}$. $A_p=\f{1}{p!}\omega_{i_1...i_p}dx^{i_1}\w...\w dx^{i_p}$ and $*A_p=\f{1}{(d-p)!}\omega_{i_1...d_p}\ep_{j_1..j_{d-p}}^{i_1...d_p}dx^{j_1}\w...\w dx^{j_{d-p}}$} 

\beqal{f2b}
S_{IIB}&=&\f{1}{2\ka^2}\left[\int d^{10}x 
\sqrt{-g}R-\f{1}{2}\int\left[d\phi\w*d\phi
+e^{-\phi}H_3\w*H_3\right]\right]\non
&-&\f{1}{4\ka_{10}^2}\left[\int e^{2\phi}F_1\w*F_1+e^{\phi} 
F_3\w*F_3+\f{1}{2}F_5\w*F_5\right]+...
\eeqa

\noi
where $\ka^2=\ka^2_{10}e^{-2\phi_0}$, and

\beqa
H_3=db \mbox{\hspace{.2in}} F_1=dC_0 \mbox{\hspace{.2in}} F_3=dC_2 
\mbox{\hspace{.2in}} F_5=dC_4
\eeqa

\noi
Where $b$ is the two form antisymmetric NS-NS field \footnote{We will denote 
the constant part of the NS-NS two form field as $B$ and the fluctuation about 
this as $b$. The same field on the brane will be identified as the field
strength of the $U(1)$ gauge field as mentioned earlier}. We have omitted the other terms in the action (\ref{f2b}) as we are only interested in the 
propagators for the closed string modes that will 
be needed to compute the two point amplitude in the later part of this 
section. The propagators for the NS-NS modes that have already
been worked out in Section~\ref{cse} in the context of bosonic string theory.

The R-R modes that will be relevant for our discussions are the zero-form, 
$C_0$ and the two-form, $C_2$.
For the R-R modes, the propagators are same as that of the NS-NS modes upto 
normalisations,

\beqal{rrp}
\expt{C_0C_0}&=&\f{\ka_{10}^2}{\ka^2}\expt{\phi\phi}\non
\expt{C_{2IJ}C_{2I^{'}J^{'}}}&=&\f{\ka_{10}^2}{\ka^2}
(2\pi\al)^2\expt{b_{IJ}b_{I^{'}J{'}}}
\eeqa

The propagators will be restricted to the 
values for the closed string metric $g^{ij}$ in the various limits stated 
at thebegining of this section. The correction to the quadratic term due to the tree-level closed string exchanges will be obtained as shown in eqn(\ref{eff}). 

\subsubsection{NS-NS exchange}\label{fns}

We now turn to the DBI and 
the Chern-Simons action of a $D_3$ brane for calculating the 
massless closed string couplings to the gauge field
The NS-NS field content is same as that of the bosonic theory. To compute 
the amplitude due to the exchange of these fields we need to set 
$D\rightarrow 10$ in the amplitudes calculated in Sections~\ref{cse1}, 
\ref{cse2} and \ref{cse3} in equations
(\ref{final1}),(\ref{final2}) and (\ref{final3}) respectively.

\subsubsection{R-R exchange}\label{frr}
The couplings of the R-R modes to the gauge field on the brane 
will be given by the usual Chern-Simons terms. We will consider here 
the commutative description of these terms. For a discussion of noncommutative
description see \cite{mukhi,liu}

\beqal{fcs}
S_{CS}=i\mu_3\int_4\sum_n C_n\w e^{2\pi\al(B+b)}
\eeqa

\noi
Expanding (\ref{fcs}) and picking out the forms proportional to the volume 
form with one $b$ insertion we get

\beqal{fcse}
S_{CS}=i\mu_3\left[(2\pi\al)^2\int_4 C_0 B\w b + (2\pi\al)\int_4 C_2\w b 
\right]
\eeqa

\begin{figure}[t]
\begin{center}
\begin{psfrags}
\psfrag{B}[][]{\footnotesize{$B$}}
\psfrag{F}[][]{\footnotesize{$F$}}
\psfrag{c2}[][]{\footnotesize{$C_2$}}
\psfrag{c0}[][]{\footnotesize{$C_0$}}
\psfrag{i}[][]{\footnotesize{(i)}}
\psfrag{ii}[][]{\footnotesize{(ii)}}
\epsfig{file=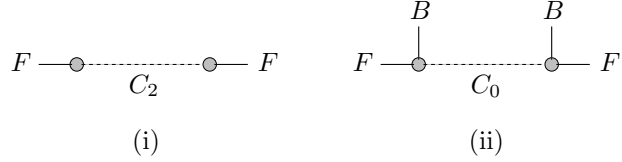, width= 8 cm,angle=0}
\end{psfrags}
\vspace{ .1 in }
\caption{Two point amplitude upto quadratic order in $B$. (i) is due to 
$C_2$ exchange (ii) due to $C_0$ exchange.}
\label{rr1}
\end{center}
\end{figure}

\noi
\underline{{\bf\it $C_2$ Exchange}} :
\\
\\
\noi
The coupling of $C_2$ to $b$ is given by,

\beqal{vc21}
V_{bC_2}=\f{i\mu_3}{4}(2\pi\al)\ep^{ijkl}
\eeqa

The two point amplitude can be worked out as in (\ref{eff}). 
For the noncommutative cases, we will rewrite the coupling (\ref{vc21}) as 

\beqal{vc22}
\lefteqn{V_{bC_2}=\sqrt{2\pi\al 
B}\f{i\mu_3}{32(2\pi\al)}\left(\f{1}{B}\right)^{pq}
\left(\f{1}{B}\right)^{rs} \ep_{pqrs}\ep^{ijkl}}\\
& &= \sqrt{2\pi\al B}\f{i\mu_3}{4(2\pi\al)}\left[\left(\f{1}{B}\right)^{ik}
\left(\f{1}{B}\right)^{jl}-\left(\f{1}{B}\right)^{jk}\left(\f{1}{B}\right)^{il}-
\left(\f{1}{B}\right)^{ij}\left(\f{1}{B}\right)^{kl}\right]\nn
\eeqa
\\
\\
Where we have used the fact that, for an 
antisymmetric matrix $M$ of rank $2n$, 
$\sqrt{M}=\f{(-1)^n}{2^n n!}\ep_{\mu_1..\mu_{2n}}M^{\mu_1\mu_2}...M^{\mu_{2n-1}
\mu_{2n}}$. The two point function can now be calculated with the above 
vertices. Note that the dependence of the amplitude on the closed string
metric $g$ only comes from the propagator. 

\
\begin{figure}[t]
\begin{center}
\begin{psfrags}
\psfrag{B1}[][]{\footnotesize{$1/B$}}
\psfrag{B2}[][]{\footnotesize{$1/B^2$}}
\psfrag{F}[][]{\footnotesize{$F$}}
\psfrag{c2}[][]{\footnotesize{$C_2$}}
\psfrag{c0}[][]{\footnotesize{$C_0$}}
\psfrag{i}[][]{\footnotesize{(i)}}
\psfrag{ii}[][]{\footnotesize{(ii)}}
\epsfig{file=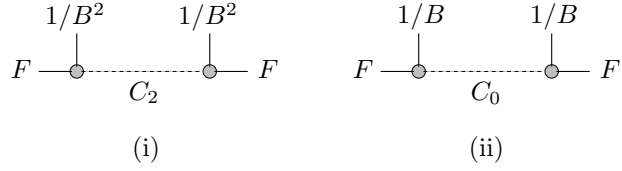, width= 8 cm,angle=0}
\end{psfrags}
\vspace{ .1 in }
\caption{Two point amplitude upto ${\cal O}(\ep^2)$ and ${\cal O}(\alpha^{'2})$  . (i) is due to 
$C_2$ exchange (ii) due to $C_0$ exchange.}
\label{rr2}
\end{center}
\end{figure}

\noi
\underline{{\it $C_0$ Exchange}} :\\
\\
\noi
Reading from (\ref{fcse}) the coupling of $C_0$ to the gauge field on the brane is given by

\beqa
V_{bC_0}=\f{i\mu_3}{4}(2\pi\al)^2 B_{ij}\ep^{ijkl}
\eeqa

\noi
The noncommutative couplings can be obtained as the $C_2$ case, and 
finally contracting the answer with $B_{ij}$. This gives

\beqa
V_{bC_0}=\sqrt{2\pi\al B}\f{i\mu_3}{2}\left(\f{1}{B}\right)^{kl}
\eeqa

\noi
The $C_2$ and $C_0$ exchanges have been summarised for various limits of $g$ 
in Figures~\ref{rr1} 
and \ref{rr2}. For details of calculation see \cite{myopcl3}.
One can easily show that, with the identification $\ka_{10}\mu_3=\ka\tau_3$, 
the full contribution to the two point function including 
both the massless NS-NS exchanges in Section~\ref{cse}, and R-R exchanges 
considered here, vanishes. We know from the one loop string calculation that 
the one loop two point amplitude vanishes for the flat space background, 
(\ref{ssum}) . In the closed string picture, this cancellation takes place 
for every mass-level between the NS-NS and R-R states. We will consider similar exchanges for the massless closed strings on the $C^2/Z_2$ orbifold in the next section.

\subsection{Type IIB on $C^2/Z_2$ orbifold}\label{orbifold}

In this section we analyse the closed string exchanges on the orbifold that 
we are ultimately interested in. 
The procedure followed is same as that of the earlier section. We will first 
write down the supergravity action on the $C^2/Z_2$ orbifold. We then derive 
the couplings of the massless closed string modes to the gauge field on a 
fractional $D_3$. We shall primarily make use of the fact that the $Z_2$ 
orbifold is the singular limit of a smooth ALE space known as Eguchi-Hanson 
space (See Appendix~\ref{ehspace}). 
The orbifold singulatity arises as the radius of the compact 2-sphere reduces 
to zero size. The compact 2-sphere (${\cal C}_1$) has an associated 
antiself-dual two form, $\omega_2$ (\ref{omega2}) that is dual to ${\cal C}_1$ 
and satisfies,

\beqal{conv}
\omega_2=-*\omega_2 \mbox{\hspace{.2in}} \int_{{\cal C}_1}\omega_2=1 
\mbox{\hspace{.2in}} \int_{C^2/Z_2}*\omega_2\w \omega_2=\f{1}{2}
\eeqa

\noi
Although the cycle ${\cal C}_1$ shrinks to zero size, there is a non-zero two-form flux $\hat{B}$. A ($p+2$)-form may be dimensionally reduced as follows,
$A_{p+2}=\tilde{A}_p\w \omega_2$. Where $\tilde{A}_p$ is a $p$-form in the transverse six dimensions. This field is twisted and is localised at the orbifold point.
For the problem at hand we also turn on the ($B+b$) field along the non-orbifolded directions, so that the background is given by

\beqal{matrix}
{\cal B}=
\left(\begin{tabular}{cccc|ccl}
0&1&2&3&...&8&9\\
&&&&&\\
&&$2\pi\al(B+b)$&&&&\\
&&&&&&\\
\hline
&&&&&&\\
&&&&&&$\hat{B}$\\
\end{tabular}\right)
\eeqa

\noi
With the above observations and using equations (\ref{conv}), we can now write down the supergravity action on the orbifold for the twisted fields,

\beqal{actorb}
S_{orb}=-\f{1}{8\ka^2}\int_6 d\tilde{b}\w * d\tilde{b}-\f{1}{8\ka_{10}^2} \int_6 \left[ d\tilde{C_0}\w *d\tilde{C_0}+d\tilde{C_2}\w *d\tilde{C_2}\right]
\eeqa

\noi
Where $\tilde{b}$ is the twisted NS-NS scalar that arises from the dimensional reduction of the $\hat{B}$ so that $\hat{B}=\hat{b}\omega_2$ and,

\beqa
\hat{b}
= 4\pi^2\al\left(\f{1}{2}+\f{\tilde{b}}{4\pi^2\al}\right)
\eeqa

\noi
$\tilde{b}$ is the fluctuating part of $\hat{b}$. Similarly the scalar, $\tilde{C}_0$ and the two-form field $\tilde{C}_2$ arises from the dimensional reduction of the R-R fields $C_2$ and $C_4$ respectively. The propagators for these twisted fields can be easily read off from (\ref{actorb}),

\beqal{ptb}
\expt{\tilde{b}\tilde{b}}=-4i\ka^2\f{1}{\kpe^2+g^{ij}\kpai\kpaj} 
\eeqa

\beqal{ptc0}
\expt{\tilde{C_0}\tilde{C_0}}-4i\ka_{10}^2\f{1}{\kpe^2+g^{ij}\kpai\kpaj} 
\eeqa

\beqal{ptc2}
\expt{\tilde{C_2}\tilde{C_2}}=-4i\ka_{10}^2\f{g_{I[J{'}}g_{I^{'}]J}}
{\kpe^2+g^{ij}\kpai\kpaj}
\eeqa

\subsubsection{NS-NS exchange}\label{ons}

For the NS-NS twisted sector we have only the $\tilde{b}$ field arising 
 from the dimensional reduction of the two form field $\hat{B}$.
In the picture outlined in the begining of Section~\ref{orbifold}, we can view a fractional $D_p$ brane as $D_{p+2}$ brane wrapped on the shrinking cycle ${\cal C}_1$. The Born-Infeld action for a $D_{p+2}$ is 

\beqal{odbi}
S_{p+2}&=&-\tau_{p+2}\int d^{p+3}\xi e^{\f{p-1}{4}\phi}\sqrt{g+{\cal B}e^{-\f{\phi}{2}}}
\eeqa

\noi
where ${\cal B}$ is given by (\ref{matrix}). We can rewrite (\ref{odbi}) as,

\beqal{odbi1}
S_p&=&-\tau_{p+2}\int d^{p+1}\xi e^{\f{p-3}{4}\phi}\sqrt{g+2\pi\al(B+b)e^{-\f{\phi}{2}}}
\int d^2\xi_{\mbox{int}}\sqrt{\hat{B}}\non
&=&-\tau_p\int d^{p+1}\xi e^{\f{p-3}{4}\phi}\sqrt{g+2\pi\al(B+b)e^{-\f{\phi}{2}}}
\left(\f{1}{2}+\f{\tilde{b}}{4\pi^2\al}\right)
\eeqa

\noi
where

\beqal{redb}
\int d^2\xi_{\mbox{int}}\sqrt{\hat{B}}=\int_{{\cal C}_1} \hat{B}
= 4\pi^2\al\left(\f{1}{2}+\f{\tilde{b}}{4\pi^2\al}\right)
\eeqa

\noi
In the second line of (\ref{odbi1}) we have identified,

\beqa
\tau_p=\tau_{p+2}(4\pi^2\al)
\eeqa

\noi
This action gives the coupling of the twisted field $\tilde{b}$ to the gauge field.
Note that we also have the untwisted NS-NS modes. 
For the untwisted states the couplings and the two point functions are the 
same as those computed in Section~\ref{cse} upto overall constants. 
Here we 
will only be concerned with the twisted field $\tilde{b}$, as the sum of the
untwisted exchanges vanish. 
We can write down the coupling of this field to the gauge field $b$ by 
expanding (\ref{odbi}) with various limits of $g$ and restricting ourselves
to $p=3$.\\
\\
\noi
\underline{{\it $\tilde{b}$ exchange}} :\\

\beqa
V_{b\tilde{b}}=\f{2\pi\al}{4\pi}\tau_3 B^{kl} \mbox{\hspace{.2in}(For small $B$ and $g=\eta$)}
\eeqa

\beqa
V_{b\tilde{b}}=\f{\sqrt{2\pi\al B}}{4\pi^2\al}\tau_3\left[\f{1}{2}\left(\f{1}{B}\right)^{kl}+
\f{\ep^2}{2(2\pi\al)^2}\left[\left(\f{1}{B^3}\right)^{kl}-\f{1}{4}\left(\f{1}{B}\right)^{kl}\Tr\left(\f{1}{B^2}\right)\right]\right]\non
\mbox{\hspace{.1in}(For $g=\ep\eta$)}\non
\eeqa

\beqa
V_{b\tilde{b}}=\f{\sqrt{2\pi\al B}}{4\pi^2\al}\tau_3\left[\f{1}{2}\left(\f{1}{B}\right)^{kl}+
\f{(2\pi\al)^2}{2}\left[B^{kl}-\f{1}{4}\left(\f{1}{B}\right)^{kl}\Tr\left({B^2}\right)\right]\right]\non
\mbox{\hspace{.1in}(For $G=\eta$)}\non
\eeqa

\begin{figure}[t]
\begin{center}
\begin{psfrags}
\psfrag{B}[][]{\footnotesize{$B$}}
\psfrag{F}[][]{\footnotesize{$F$}}
\psfrag{c2}[][]{\footnotesize{$\tilde{C}_2$}}
\psfrag{c0}[][]{\footnotesize{$\tilde{C}_0$}}
\psfrag{tb}[][]{\footnotesize{$\tilde{b}$}}
\psfrag{i}[][]{\footnotesize{(i)}}
\psfrag{ii}[][]{\footnotesize{(ii)}}
\psfrag{iii}[][]{\footnotesize{(iii)}}
\epsfig{file=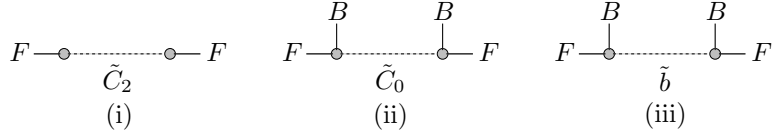, width= 10 cm,angle=0}
\end{psfrags}
\vspace{ .1 in }
\caption{Two point amplitude upto quadratic order in $B$. (i) is due to 
$\tilde{C}_2$ exchange (ii) due to $\tilde{C}_0$ exchange and (iii) due to $\tilde{b}$ exchange.}
\label{rr3}
\end{center}
\end{figure}

\subsubsection{R-R exchange}\label{orr}

As in Section~\ref{ons}, we now expand the 
Chern-Simons action in terms of the twisted and the untwisted R-R fields. 
Keeping in mind the background two-form field (\ref{matrix}) and the 
relations (\ref{conv}). We start with the action for a $D_5$ brane 
wrapping a two cycle ${\cal C}_1$. 

\beqal{ocs}
S_{CS}&=&i\mu_5\int_{6}\sum_nC_n\w e^{{\cal B}}\non
&=&
i\mu_3\f{1}{2}\left[(2\pi\al)^2\int_4 C_0 B\w b + (2\pi\al)\int_4 C_2\w b\right]\non
&+&i\mu_3\f{1}{4\pi^2\al}\left[(2\pi\al)^2\int_4 \tilde{C}_0 B\w b + (2\pi\al)\int_4 \tilde{C}_2\w b \right]
\eeqa

\noi
Where in the second line of (\ref{ocs}) we have identified $\mu_3=\mu_5(4\pi^2\al)$.
Note that the twisted and untwisted R-R couplings are same as those computed in Section~\ref{frr}  except for the change in the overall normalisations.
The R-R exchanges for the twisted states thus have the same tensor structures 
as those in Section~\ref{frr}. Incorporating these changes the two point function with twisted R-R exchanges can easily be computed.

\begin{figure}[t]
\begin{center}
\begin{psfrags}
\psfrag{B1}[][]{\footnotesize{$1/B$}}
\psfrag{B2}[][]{\footnotesize{$1/B^2$}}
\psfrag{B3}[][]{\footnotesize{$1/B^3$}}
\psfrag{F}[][]{\footnotesize{$F$}}
\psfrag{c2}[][]{\footnotesize{$\tilde{C}_2$}}
\psfrag{c0}[][]{\footnotesize{$\tilde{C}_0$}}
\psfrag{tb}[][]{\footnotesize{$\tilde{b}$}}
\psfrag{i}[][]{\footnotesize{(i)}}
\psfrag{ii}[][]{\footnotesize{(ii)}}
\psfrag{iii}[][]{\footnotesize{(iii)}}
\epsfig{file=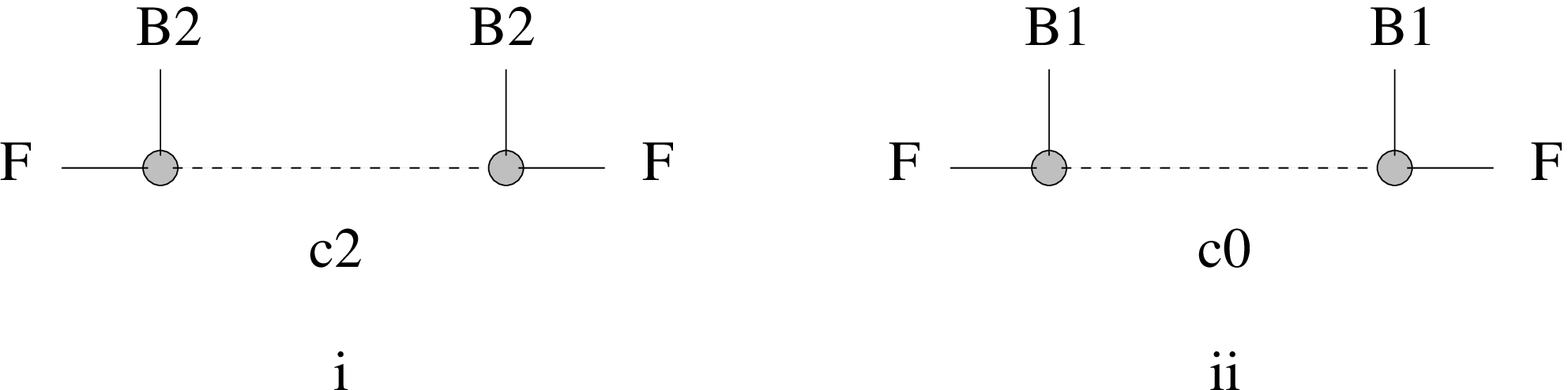, width= 10 cm,angle=0}
\end{psfrags}
\vspace{ .1 in }
\caption{Two point amplitude upto ${\cal O}(\ep^2)$. (i) is due to 
$\tilde{C}_2$ exchange (ii) due to $\tilde{C}_0$ exchange and (iii) due to $\tilde{b}$ exchange.}
\label{rr4}
\end{center}
\end{figure}

\begin{figure}[t]
\begin{center}
\begin{psfrags}
\psfrag{B1}[][]{\footnotesize{$1/B$}}
\psfrag{B2}[][]{\footnotesize{$1/B^2$}}
\psfrag{B}[][]{\footnotesize{$B$}}
\psfrag{F}[][]{\footnotesize{$F$}}
\psfrag{c2}[][]{\footnotesize{$\tilde{C}_2$}}
\psfrag{c0}[][]{\footnotesize{$\tilde{C}_0$}}
\psfrag{tb}[][]{\footnotesize{$\tilde{b}$}}
\psfrag{i}[][]{\footnotesize{(i)}}
\psfrag{ii}[][]{\footnotesize{(ii)}}
\psfrag{iii}[][]{\footnotesize{(iii)}}
\epsfig{file=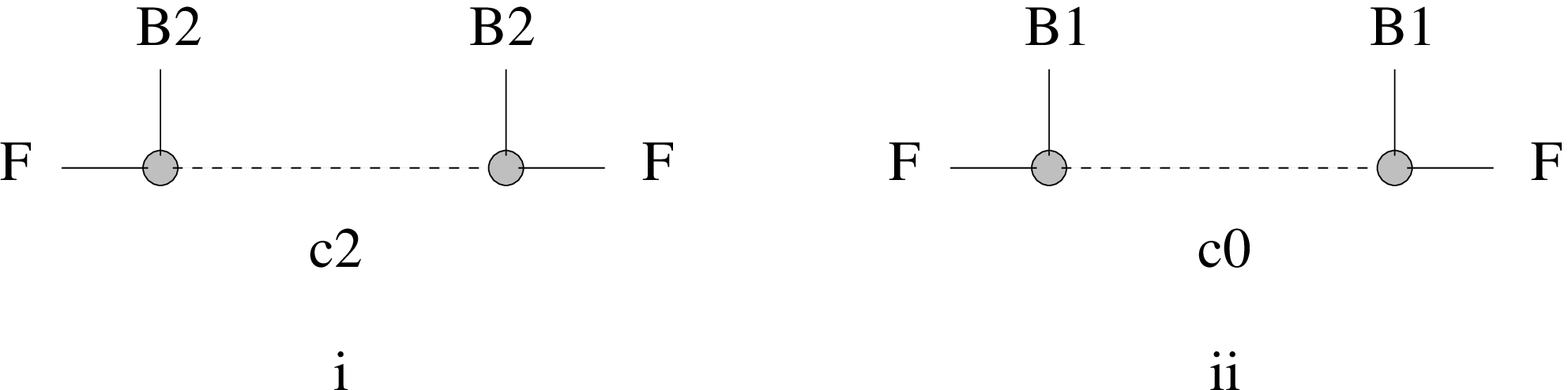, width= 10 cm,angle=0}
\end{psfrags}
\vspace{ .1 in }
\caption{Two point amplitude upto ${\cal O}(\alpha^{'2})$. (i) is due to 
$\tilde{C}_2$ exchange (ii) due to $\tilde{C}_0$ exchange and (iii) due to 
$\tilde{b}$ exchange.}
\label{rr5}
\end{center}
\end{figure}

The twisted closed string exchanges in the various limits of $g$ are
shown in Figures~\ref{rr3}, \ref{rr4} and \ref{rr5}.
We have seen that the untwisted exchanges for both the NS-NS and R-R sectors 
are the same as those computed in section (\ref{flat}) modulo an overall 
normalisation. The sum of these thus vanishes just like the flat case. 
This is also what we get from the one loop computation. 
See eqn(\ref{sum}). The twisted states from both the NS-NS and R-R sectors however sum up to finite results. We 
write these contributions below with the identification 
$\ka_{10}\mu_3=\ka\tau_3$.
\\
\\
\noi
For small $B$,

\beqa
L_2&=&\f{i}{{4\pi^2}}\ka_{10}^2\mu_3^2  \int\f{d^2\kpe}{(2\pi)^2}
\f{1}{\kpe^2+p^2}\times\non&\times&
[\f{1}{2}\left[1-(2\pi\al)^2\f{1}{2}\Tr(B^2)\right]\left(\eta^{kk^{'}}\eta^{ll^{'}}-\eta^{kl^{'}}\eta^{lk^{'}}\right)
\non &+&(2\pi\al)^2\left[\eta^{kk^{'}}(B^2)^{ll^{'}}-\eta^{kl^{'}}(B^2)^{lk^{'}}
\right]+(kl) \leftrightarrow (k^{'}l^{'})] 
\eeqa

\noi
For $g=\ep\eta$,

\beqa
L_2&=&\f{i}{{4\pi^2}}\mbox{det}(2\pi\al
B)\ka_{10}^2\mu_3^2\int
\f{d^2\kpe}{(2\pi)^2}\f{1}{\kpe^2+\ep^{-1}p^2}
\times \non
&\times&\f{\ep^2}{(2\pi\al)^4}
\f{1}{2}\left[\left(\f{1}{B^2}\right)^{kk^{'}}\left(\f{1}{B^2}\right)^{ll^{'}}
-\left(\f{1}{B^2}\right)^{k^{'}l}\left(\f{1}{B^2}\right)^{kl^{'}}\right]
\non &+& (kl) \leftrightarrow (k^{'}l^{'})
\eeqa

\noi
For $G=\eta$,

\beqa
L_2
&=&
\f{i}{{4\pi^2}}\mbox{det}(2\pi\al B)\ka_{10}^2\mu_3^2
\int \f{d^2\kpe}{(2\pi)^2}
\f{1}{\kpe^2+\tilde{p}^2/(2\pi\al)^2}\times\non &\times&
\f{1}{2}\left[\eta^{ll^{'}}\eta^{kk^{'}}
-\eta^{kl^{'}}\eta^{lk^{'}}\right]\non &+& (kl) \leftrightarrow
(k^{'}l^{'})
\eeqa

\noi
These expressions are the expansions of the one loop sting amplitude, 
eqn(\ref{sum}) when $G$ and $\det(g+2\pi\al B)$ are expanded to 
respective orders as in equations (\ref{expan1}), (\ref{expan2}) and (\ref{expan3}). We have noted earlier in the bosonic case that the 
computation of the two point amplitude in string theory sums up all 
the $B$-field dependence in the open string metric $G$ and $\det(g+2\pi\al B)$. The analysis in this section again reproduces these terms to the 
orders relevant in the expansion about the various limits of the 
closed string metric $g$.

\section{Discussions}\label{dis}

\noindent

We now discuss some of the important issues addressed in this article. 
The central theme has been the world sheet open-closed
duality in the presence of $B$-field. The primary 
motivation for studying this is to see the UV/IR correspondence in 
noncommutative gauge theories and to identify the exact role played by 
the $B$-field. World-sheet duality
underlies the duality between gravity and gauge theory. An example of 
which is the AdS/CFT conjecture \cite{adscft,nadscft}\footnote{For further studies related
to the roles played by the ordinary and noncommutative Yang-Mills theories see
\cite{Cai:1999aw}.}. On the gauge theory side we have a 
${\cal N}=4$ superconformal theory. This theory is finite and hence 
the duality of the annulus diagram between the open and the closed 
string channels reduces
to a trivial identity namely, $0=0$. For some nontrivial correspondence 
one must reduce the amount of supersymmetry and break conformal 
invariance without however bringing in tachyons.
In this case one loop amplitudes are
divergent. One can compare divergences in the closed and open string 
channels and
if one makes a suitable identification of the cutoffs one can show the
equality of amplitudes. We have seen that the $B$-field 
plays the role of a regulator for the nonplanar diagrams and preserves the 
duality. 

\begin{figure}[t]
\begin{center}
\begin{psfrags}
\psfrag{Open}[][]{{\footnotesize Open String}}
\psfrag{Loop}[][]{{\footnotesize One loop amplitude}}
\psfrag{ncfield}[][]{{\footnotesize Noncommutative}}
\psfrag{theory}[][]{{\footnotesize Field Theory}}
\psfrag{sw}[][]{{\footnotesize SW Limit}}
\psfrag{downa}[][]{{\footnotesize $\al \sim \epsilon^{1/2} \rightarrow 
0$}}
\psfrag{downaa}[][]{{\footnotesize $g_{ij} \sim \epsilon \rightarrow 
0$}}
\psfrag{downb}[][]{{\footnotesize $t \rightarrow \infty$}}
\psfrag{right}[][]{{\footnotesize $t \rightarrow 1/t$}}
\psfrag{closed}[][]{{\footnotesize Closed 
String}}
\psfrag{channel}[][]{{\footnotesize Channel}}
\psfrag{downc}[][]{{\footnotesize $t \rightarrow 0$}}
\psfrag{massless}[][]{{\footnotesize Massless Closed}}
\psfrag{exchange}[][]{{\footnotesize String Exchange}}
\psfrag{uv}[][]{{\footnotesize UV}}
\psfrag{ir}[][]{{\footnotesize IR}}
\epsfig{file=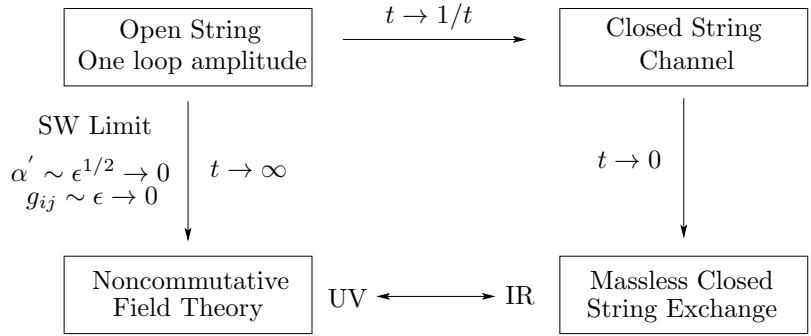, width= 10 cm,angle=0}
\end{psfrags}
\vspace{ .1 in }
\caption{Noncommutative field theory and closed string channel
limits}
\label{lim2}
\end{center}
\end{figure}

On the gauge theory side, the presence of the $B$-field leads to mixing of
UV and IR.
The annulus diagram in the Seiberg-Witten limit gives the one 
loop diagram in noncommutative gauge theory. One also needs to keep only the 
massless modes propagating in the loop that survive in the $t \rightarrow 
\infty$ limit. 
It is important to note that UV divergence in the gauge theory originates from 
the same
region of the modulus as the massless closed string exchanges i.e. $t
\rightarrow 0$ end.
In general the open string loop UV region is reproduced by 
closed string trees with small
(i.e IR) momentum exchange. A flow-chart of the analysis followed in this article is shown in Figure~\ref{lim2}.\\
\\

\noi
$\bullet$
{\it Bosonic Theory}\\
\\
\noi
In this bosonic theory, we have seen that the divergences arising from 
the two ends 
are related to each other (upto some overall normalisation). This 
relation could not be made exact in the bosonic setup due to 
the presence of tachyons, which act as additional sources for divergences.
We have shown that the tensor structure for the noncommutative 
field theory (\ref{photon2pt}) two point amplitude can be recovered by 
considering 
massless and tachyonic exchanges of closed strings in the 
presence of background constant $B$-field. 
For the coefficients to match with the gauge theory result, in the 
bosonic string case, the full tower of the closed string states are required. 
We will comment on the closed string couplings below.\\

\noi
$\bullet$
{\it Superstrings on $C^2/Z_2$}\\
\\
\noi 
With the lessons from the bosonic theory, in 
Sections~\ref{osbf} and 
\ref{tpa} we have studied a noncommutative ${\cal N}=2$ gauge theory 
realised on a fractional $D_3$-brane localised at the fixed point of $C^2/Z_2$ 
orbifold. 
As discussed above, the UV divergences of the gauge theory 
comes from the $t\rightarrow 0$ end that is dominated by massless 
closed strings. In general the massless closed 
string exchanges account for the UV contribution due to all the open 
string modes and similarly the dual description  of the gauge 
theory would thus require the contributions from all the massive closed 
string states as well. This is the situation in the bosonic case. But in this 
supersymmetric case, the contributions from the massive modes cancel and hence 
the duality is between the finite number of massless states on both the open 
and the closed string sides. This is reflected in 
the equality of (\ref{final}) to the gauge theory amplitude, with both 
ends regulated in the presence of the $B$-field.\\

\noi
$\bullet$
{\it Closed string couplings}\\
\\
\noindent
Let us now discuss about the closed string couplings to the 
noncommutative 
gauge theory on the brane. Consider for the 
moment, eqn(~\ref{opc}) in the bosonic theory. Let us set $\alpha ' t = T$
\beqal{opc2}
{\cal A}_{op} \sim \int \f{dT}{T} (T)^{-\f{p+1}{2}}
\left[e^{2\pi {T\over \alpha '}}+(D-2)+O(e^{-2\pi {T\over \alpha '}})\right]\exp(-C/T)
\eeqa

The $O(1)$ term in the expansion corresponds to the massless open string 
modes in the loop. If we take the $\alpha ' \rightarrow 0$ 
limit the contribution of the massive modes drop out. If we ignore the tachyon
we get the massless mode contribution. In the supersymmetric case there is no tachyon.
However in the present case dropping the tachyon term makes an exact comparison
of the massless sectors of the two cases meaningless because the powers of $\alpha '$
cannot match. Nevertheless the comparison is instructive.  

The UV contribution of (\ref{opc},\ref{opc2}), as shown in 
Figure~\ref{lim1}, comes from the region $0<t<1/\Lambda^2\al$. The UV 
divergences coming from this region is regulated by $C$. In the closed 
string channel we have,

\beqal{clc1}
{\cal A}_{cl} &\sim& \int ds (\al)^{-\f{p+1}{2}}s^{-l/2}
\left[e^{2\pi s}+(D-2)+O(e^{-2\pi s})\right]\exp(-Cs/\al )\non
&\sim& (\al)^{-\f{p+1}{2}} (\al)^{\f{l}{2}-1}\int 
d^l\kpe\f{1}{\kpe^2+C/\als}
\eeqa

The $\alpha '\rightarrow 0$ limit does not pull out the massless
sector (even if we ignore the tachyon) and this makes it clear
that in general all the massive closed string modes are required to 
reproduce the massless open string contribution.
But let us focus on the massless states of the closed string sector.
In the second expression of (\ref{clc1}), we have kept only the 
contribution from the massless closed string mode.
This expression can be interpreted as the amplitude of emission and 
absorption of a closed string 
state from the $D_p$-brane with transverse momentum  $\kpe$, integrated 
over $0<\kpe<1/\Lambda\al$. The domain of the $\kpe$ integral 
corresponds to the UV region in the open string channel.   
$l=D-(p+1)$, is the number of transverse directions in which the 
closed string propagates. 
For $l\ne 2$ there is an extra factor of $(\al)^{\f{l}{2}-1}$
in (\ref{clc1}) over (\ref{opc}), that makes the couplings of the 
individual closed string modes vanish when compared to the open string 
channel. It is only for the special case $l=2$ that the powers match.

In the supersymmetric case,
for the $C^2/Z_2$ orbifold, we have seen that the closed strings that 
contribute to the dual
description of the nonplanar divergences are from the twisted sectors.
They are free to move in $6$ directions transverse to the orbifold. For 
the $D_3$-brane that is localised at the fixed point with world volume 
directions perpendicular to the orbifold, 
these closed string twisted states propagate in 
exactly two directions transverse to the brane. Thus in this case $l=2$ 
and from the above 
discussions this makes the power of $\al$ in the coupling of closed 
string 
with the gauge field strength same as that of the open string channel.

In general, the closed string couplings to the gauge field when 
the closed string modes are restricted to the massless ones do not 
give the same normalisation as the gauge theory. The massive closed 
string modes are expected to contribute so that the normalisations at 
both the ends are equal. This is guaranteed by open-closed string 
duality. For the $C^2/Z_2$ orbifold, the massive states cancel. 
The finite number of closed string modes thus give the same 
normalisation as the gauge theory two point function. This is the reason 
why we are able to see the IR behaviour of noncommutative ${\cal N}=2$ 
theory in terms of only the massless closed string modes in the twisted 
sectors.\\

\noi
$\bullet$
{\it Other models}\\
\\
\noi
We conclude this article with some comments about other models where this UV/IR 
correspondence may be seen exactly. In the gauge theory we are able to see
the mixing of the UV and IR sectors due to the nature of ultraviolet regulation 
by the $B$-field. However we have seen that apart from the regulatory nature
of the $B$-field, its contribution to the partition function factors out in 
the determinant (\ref{det}). This means that the $B$-field does not disturb the
duality between the massless open/closed modes that may be present in the
commutative models. UV/IR correspondence between the noncommutative gauge theory and gravity will thus be naturally manifested in these models. We have seen
how this works for the $C^2/Z_2$ orbifold. Another example is the $C^3/(Z_2\times Z_2)$ orbifold that gives ${\cal N}=1$ gauge theory on the fractional $D_3$ brane. One may ask whether other such models exist. An analysis on this issue have
been made in \cite{DiVecchia:2005vm}. Finally, in this article we have only addressed the 
perturbative nature of this UV/IR correspondence. It will be interesting to explore
the non-perturbative aspects of this duality possibly along the lines of \cite{Das:1999kg,Billo:2005fg}. \\

\noindent
{\bf Acknowledgements}

\noindent
This article is based on works with B. Sathiapalan.
I would like to thank him for various helpful discussions and for 
carefully reading this manuscript. 
I would also like to thank the members of the String Theory groups at 
University of Roma Tor Vergata and University of Torino for useful discussions.

\appendix
\section{Vacuum amplitude}\label{vacuum}

In this appendix we calculate the vacuum amplitude for the open strings 
with end points on a $D_3$-brane that is located at the fixed point of 
$C^2/Z_2$ orbifold. Let us first start with the bosonic part of the 
world-sheet action, 

\beqa
S_B=-\f{1}{4\pi\al}\int_{\Sigma}g_{MN}\pa_a X^M\pa^{a}X^N
+\f{1}{2}\int_{\pa\Sigma}B_{MN}X^M\pa_{\tau}X^N\\
\eeqa

The boundary condition for the world-sheet bosons from the above action 
is,

\beqal{bcb}
g_{MN}\pa_{\s}X^N+2\pi\al B_{MN}\pa_{\tau}X^N=0\mid_{\s=0,\pi}
\eeqa

In the Seiberg-Witten limit, $g_{ij}=\ep\eta_{ij}$ we choose the $B$
field along the brane to be of the form,

\beqal{bmatrix}
B=\f{\ep}{2\pi\al}
\left( \begin{array}{cccc}
0 & b_1 & 0 & 0\\
-b_1 & 0 & 0 &0 \\
0 & 0 & 0 & b_2 \\
0& 0& -b_2 & 0
\end{array} \right)
\eeqa

With the above form for the $B$-field, and defining,

\beqa
X_{(1)}^{\pm}=2^{-1/2}(X^0\pm X^1)
\mbox{\hspace{0.2in}and\hspace{0.2in}}
X_{(2)}^{\pm}=2^{-1/2}(X^2\pm iX^3)
\eeqa

the boundary condition (\ref{bcb}) can be rewritten as,

\beqa
\pa_{\s}X_{(1)}^{\pm}=\pm b_1\pa_{\tau}X_{(1)}^{\pm}\mid_{\s=0,\pi}
\mbox{\hspace{0.2in}and\hspace{0.2in}}
\pa_{\s}X_{(2)}^{\pm}=\pm ib_2\pa_{\tau}X_{(1)}^{\pm}\mid_{\s=0,\pi}
\eeqa

The mode expansions for the open string satisfying the above boundary 
conditions are given by,

\beqal{mb}
X^{\pm}_{(1)}&=&x^{\pm}_{(1)}+\f{2\al}{1-b_1^2}
(\tau\pm b_1\s)p^{\pm}_{(1)}
+i\sqrt{2\al}\sum_{n\neq 0}\f{a^{\pm}_{(1)n}}{n}e^{-i(n\tau \pm \nu_1)}
\cos(n\s \mp \nu_1)\non
X^{\pm}_{(2)}&=&x^{\pm}_{(2)}+\f{2\al}{1+b_2^2}
(\tau\pm ib_2\s)p^{\pm}_{(2)}
+i\sqrt{2\al}\sum_{n\neq 0}\f{a^{\pm}_{(2)n}}{n}e^{-i(n\tau \pm \nu_2)}
\cos(n\s \mp \nu_2)\non
\eeqa

\noindent
where we have defined,

\beqa
i\nu_1=\f{1}{2}\log\left(\f{1+b_1}{1-b_1}\right) \mbox{\hspace{0.5in}}
i\nu_2=\f{1}{2}\log\left(\f{1+ib_2}{1-ib_2}\right)
\eeqa

\noindent
The coefficients of the mode expansions (\ref{mb}) are fixed so as to 
satisfy,

\beqa
\left[X^{+}_{(1)}(\tau,\s),P^{-}_{(1)}(\tau,\s^{'})\right]=-2\pi\al\de(\s-\s^{'})
\eeqa

\noindent
and that the  zero modes and the other oscillators satisfy the usual 
commutation relations,

\beqa
\left[a^{+}_{(1)m},a^{-}_{(1)n}\right]=-m\de_{m+n}
\mbox{\hspace{0.5in}}
\left[a^{+}_{(2)m},a^{-}_{(2)n}\right]=m\de_{m+n}
\eeqa

\beqa
\left[x^{+}_{(1)},p^{-}_{(1)}\right]=-i \mbox{\hspace{0.5in}}
\left[x^{+}_{(2)},p^{-}_{(2)}\right]=i
\eeqa

There is no shift in the moding of the oscillators, the 
zero point energy and the spectrum is the same as the $B=0$ case. The 
situation is the 
same as that of a neutral string in electromagnetic background 
\cite{callan}. Note 
that the commutator for $X^{\pm}$ now does not vanish at the 
boundary, for example,

\beqa
\left[X^{+}_{(1)}(\tau,0),X^{-}_{(1)}(\tau,0)\right]=-2\pi i\al 
\f{b_1}{1-b_1^2}\non
\left[X^{+}_{(1)}(\tau,\pi),X^{-}_{(1)}(\tau,\pi)\right]=2\pi i\al 
\f{b_1}{1-b_1^2}
\eeqa

The zero mode for the energy momentum tensor can now be worked out and 
is given by,

\beqa
L_{(b)0}^{\parallel}=\f{2\al}{b_1^2-1}p^{+}_{(1)}p^{-}_{(1)}
+\f{2\al}{b_2^2+1}p^{+}_{(2)}p^{-}_{(2)}
-\sum_{n\neq 0}
\left[a^{+}_{(1)-n}a^{-}_{(1)}-a^{+}_{(2)-n}a^{-}_{(2)}
\right]
\eeqa

\noindent
Since the spectrum remains the same, the contribution to the vacuum 
amplitude from the bosonic modes is the same as the usual $B=0$ case 
except that there is a factor of 
$\sqrt{(b_i^2\pm 1)}$ which comes from the trace over the zero modes for 
each direction along the brane. From (\ref{bmatrix}) in 
the limit (\ref{swl}), $b_i \sim 1/\sqrt{\ep}$ for $B$ to be finite. 
With this,

\beqal{det}
\ep^2\prod_{i}^2(b_i^2 \pm 1) \rightarrow \det(g+2\pi\al B)
\eeqa

\noindent
Including contributions from all the directions,

\beqa
L_{(b)0}=L_{(b)0}^{\parallel}+L_{(b)0}^{\perp}+L_{(b)0}^{orb}-\f{5}{12}
\eeqa

\noindent
$\perp$ denotes the $4,5$ directions and $6,7,8,9$ are the orbifolded 
directions. Let us now compute the contributions from the world sheet 
fermions. The action is given by,

\beqa
S_F=\f{i}{4\pi\al}\int_{\Sigma}g_{MN}\bps^{M}\rho^{\alpha}\pa_{\alpha}
\ps^{N}-\f{i}{4}\int_{\pa\Sigma}B_{MN}\bps^{N}\rho^0\ps^{M}
\eeqa

We rewrite the boundary equations from (\ref{bcf}),

\beqal
g_{MN}(\ps_L^N-\ps_R^N)+2\pi\al B_{MN}(\ps_L^N+\ps_R^N)=0
\mid_{\s=\pi}\\
g_{MN}(\ps_L^M+(-1)^a\ps_R^M) + 2\pi\al B_{MN}(\ps_L^N-(-1)^a\ps_R^N)=0
\mid_{\s=0}
\eeqa

Now defining,

\beqa
\ps_{(1)R,L}^{\pm}=2^{-1/2}(\ps_{R,L}^0 \pm \ps_{R,L}^1)
\mbox{\hspace{0.2in}and\hspace{0.2in}}
\ps_{(2)R,L}^{\pm}=2^{-1/2}(\ps_{R,L}^2 \pm i\ps_{R,L}^3)
\eeqa

For the Ramond Sector $(a=1)$ with the constant $B$-field given by 
(\ref{bmatrix}),

\beqa
\ps_{(1)R}^{\pm}(1\pm b_1)=\ps_{(1)L}^{\pm}(1 \mp b_1)\mid_{\s=0,\pi}
\eeqa

Mode expansion,

\beqal{modeferm1}
\ps_{(1)L,R}^{\pm}=\sum_n d_{(1)n}^{\pm}\chi_{(1)L,R}^{\pm}(\s,\tau,n)
\eeqa

where,
\beqa
\chi_{(1)R}^{\pm}=\sqrt{2\al}\exp\{-in(\tau-\s)\mp\nu_1\}\\
\chi_{(1)L}^{\pm}=\sqrt{2\al}\exp\{-in(\tau+\s)\pm\nu_1\}
\eeqa

and
\beqa
\nu_1=\f{1}{2}\log\left(\f{1+b_1}{1-b_1}\right)=\tanh^{-1}b_1
\eeqa

The boundary condition for the other two directions are,

\beqa
\ps_{(2)R}^{\pm}(1\pm ib_2)=\ps_{(2)L}^{\pm}(1 \mp ib_2)\mid_{\s=0,\pi}
\eeqa

This gives the same mode expansion as (\ref{modeferm1}),

\beqal{modeferm2}
\ps_{(2)L,R}^{\pm}=\sum_n d_{(2)n}^{\pm}\chi_{(1)L,R}^{\pm}(\s,\tau,n)
\eeqa

\beqa
\chi_{(2)R}^{\pm}=\sqrt{2\al}\exp\{-in(\tau-\s)\mp\nu_2\}\\
\chi_{(2)L}^{\pm}=\sqrt{2\al}\exp\{-in(\tau+\s)\pm\nu_2\}
\eeqa

and
\beqa
\nu_2=\f{1}{2}\log\left(\f{1+ib_2}{1-ib_2}\right)=\tan^{-1}b_2
\eeqa

\noindent
Like the bosonic partners there is no shift in the frequencies. The 
oscillators are
integer moded as usual. For the Neveu-Schwarz sector, $(a=0)$, the 
relative sign between
$\ps_{R}^{\pm}$ and $\ps_{L}^{\pm}$ at the $\s=\pi$ end in eqn(6) can be 
brought about by the 
usual restriction on $n$ to only run over half integers in the mode
expansions (\ref{modeferm1},\ref{modeferm2}).
The oscillators satisfy the standard anticommutation relations,

\beqa
\{d_{(1)n}^{+},d_{(1)m}^{-}\}=-\de_{m+n}
\mbox{\hspace{0.2in};\hspace{0.2in}}
\{d_{(2)n}^{+},d_{(2)m}^{-}\}=\de_{m+n}
\eeqa

\noindent
The zero mode for the energy momentum tensor for the fermions along 
the brane can be written as,

\beqa
L_{(f)0}^{\parallel}=\sum_n 
n\left[d_{(2)-n}^{-}d_{(2)n}^{+}-d_{(1)-n}^{-}d_{(1)n}^{+}\right] 
\eeqa

\noindent
For all the fermions including the contributions from the other 
directions we have,

\beqa
L_{(f)0}=L_{(f)0}^{\parallel}+L_{(f)0}^{\perp}+L_{(f)0}^{orb} +c_f(a)
\eeqa

\noindent
where $L_{(f)0}^{\perp}$ and $L_{(f)0}^{orb}$ have the usual 
representation in terms of oscillators.

\beqa
c_f(1)=\f{5}{12} \mbox{\hspace{0.2in};\hspace{0.2in}} c_f(0)=-\f{5}{24}
\eeqa

\noindent
We now compute the vacuum amplitude including the contributions from the 
ghosts. This is given by,

\beqa
Z_C={\cal V}_4\det(g+2\pi\al B)\int_0^{\infty}
\f{dt}{t}(8\pi^2\al t)^{-2}\Tr_{NS-R}\left[\left(\f{1+g}{2}\right)
\left(\f{1+(-1)^F}{2}\right)q^{L_0}\right]\non
\eeqa

\noindent
The origin of the $\det(g+2\pi\al B)$ term is given in (\ref{det}) and 
${\cal V}_4$ is the volume of the $D_3$-brane and $q=e^{-2\pi t}$.
The trace is summed over the spin structures with the orbifold 
projection. The required traces are listed below in terms of the {\it 
Theta Functions}, $\vth_i(\nu,it)$.

\beqal{part}
Z\zz_{e}(it)&=&\Tr_{NS}\left[q^{L_0}\right]\non&=&
\left[q^{-1/3}\prod_{m=1}^{\infty}(1-q^m)^{-8}\right]
\left[q^{-1/6}\prod_{m=1}^{\infty}(1+q^{m-1/2})^{-8}\right]\non
&=&\eta(it)^{-12}\vth^4_{3}(0,it)
\eeqa

\beqa
Z\zo_e(it)&=&\Tr_{NS}\left[(-1)^F q^{L_0}\right]\non&=&
-\left[q^{-1/3}\prod_{m=1}^{\infty}(1-q^m)^{-8}\right]
\left[q^{-1/6}\prod_{m=1}^{\infty}(1-q^{m-1/2})^{-8}\right]\non
&=&-\eta(it)^{-12}\vth^4_{4}(0,it)
\eeqa

\beqa
Z\oz_{e}(it)&=&\Tr_{R}\left[q^{L_0}\right]\non&=&
-\left[q^{-1/3}\prod_{m=1}^{\infty}(1-q^m)^{-8}\right]
\left[q^{1/3}\prod_{m=1}^{\infty}(1+q^{m})^{8}\right]\non
&=&-\eta(it)^{-12}\vth^4_{2}(0,it)
\eeqa

\beqa
Z\oo_e(it)&=&\Tr_{R}\left[(-1)^F q^{L_0}\right]=0
\eeqa

\beqa
\lefteqn{Z\zz_{g}(it) =\Tr_{NS}\left[gq^{L_0}\right] }\non 
& & =\left[q^{-1/3}\prod_{m=1}^{\infty}(1-q^m)^{-4(1+q^m)^{-4}}\right]
\left[q^{1/3}\prod_{m=1}^{\infty}(1+q^{m-1/2})^{4}(1-q^{m-1/2})^{4}
\right]\non
& &=4\eta(it)^{-6}\vth^2_{3}(0,it)\vth^2_{4}(0,it)\vth^{-2}_{2}(0,it)
\eeqa

\beqa
\lefteqn{Z\zo_g(it) =\Tr_{NS}\left[g(-1)^F q^{L_0}\right] }\non
& &=-\left[q^{-1/3}\prod_{m=1}^{\infty}(1-q^m)^{-4(1+q^m)^{-4}}\right]
\left[q^{1/3}\prod_{m=1}^{\infty}(1-q^{m-1/2})^{4}(1+q^{m-1/2})^{4}
\right]\non
& &=-4\eta(it)^{-6}\vth^2_{3}(0,it)\vth^2_{4}(0,it)\vth^{-2}_{2}(0,it)
\eeqa

\beqa
Z\oz_{g}(it)&=&\Tr_{R}\left[gq^{L_0}\right]=0 
\eeqa

\beqa
Z\oo_g(it)&=&\Tr_{R}\left[g(-1)^F q^{L_0}\right]=0\non
\eeqa

Recalling,

\beqa
\vth^4_{3}(0,it)-\vth^4_{4}(0,it)-\vth^4_{2}(0,it)=0
\eeqa

and noting that,

\beqa
Z\zz_g(it)=-Z\zo_g(it) 
\eeqa

the vacuum amplitude vanishes. This is as a result of supersymmetry.
\\
\\
\noindent
{\bf Theta Functions :}

\beqa
q=\exp(-2\pi t) \mbox{\hspace{0.5in}} z=\exp(2\pi i \nu)
\eeqa

\beqa
\vth_{00}(\nu,it)=\vth_3(\nu,it)=\prod_{m=1}^{\infty}(1-q^m)(1+zq^{m-1/2})
(1+z^{-1}q^{m-1/2})
\eeqa

\beqa
\vth_{01}(\nu,it)=\vth_4(\nu,it)=\prod_{m=1}^{\infty}(1-q^m)(1-zq^{m-1/2})
(1-z^{-1}q^{m-1/2})
\eeqa

\beqa
\vth_{10}(\nu,it)&=&\vth_2(\nu,it)=2\exp(-\pi 
t/4)\cos(\pi\nu)\times\non
&\times&\prod_{m=1}^{\infty}(1-q^m)(1+zq^{m-1/2})
(1+z^{-1}q^{m-1/2})
\eeqa

\beqa
\vth_{11}(\nu,it)&=&\vth_1(\nu,it)=-2\exp(-\pi
t/4)\sin(\pi\nu)\times\non 
&\times&\prod_{m=1}^{\infty}(1-q^m)(1-zq^{m-1/2})
(1-z^{-1}q^{m-1/2})
\eeqa

\beqa
\eta(it)=q^{1/24}\prod_{m=1}^{\infty}(1-q^m)
\eeqa

\section{Eguchi-Hanson Space}\label{ehspace}

In this appendix we give some of the properties of the Eguchi-Hanson space
\cite{Eguchi:1978gw} that are used in the analysis in Section~\ref{orbifold}.
This space is an euclidean solution of vacuum Einstein equation and is asymptotically locally euclidean. The metric is given by,

\beqal{eh2}
ds^2=f(r)^{-1}dr^2+r^2f(r)\s_z^2+r^2\left[\s_x^2+\s_y^2\right]
\eeqa

\noi where,

\beqa
f(r)=\left[1-\left(\f{a}{r}\right)^4\right] \mbox{\hspace{0.1in}and\hspace{0.1in}} \s_x=-\f{1}{2}\left(\cos{\psi}d\theta+\sin{\theta}\sin{\psi}d\phi\right)\non
\s_y=\f{1}{2}\left(\sin{\psi}d\theta-\sin{\theta}\cos{\psi}d\phi\right)
\mbox{\hspace{0.1in}} \s_z=-\f{1}{2}\left(d\psi+\cos{\theta}d\phi\right) \non
\eeqa

\noi
There is an apparent singularity at $r=a$ which is removed if one identifies
the range of $\psi$ to be $0\le\psi\le 2\pi$. $\theta$ and $\phi$ have the ranges,
$0\le \theta \le \pi$ and $0\le \phi \le 2\pi$.
The space near $r=a$ is locally
$R^2\times S^2$. This is seen from the change variables to $u^2=r^2\left[1-\left(\f{a}{r}\right)^4\right]$, so that for $r=a$ or $u=0$

\beqa
ds^2 \sim \f{1}{4}du^2+\f{1}{4}u^2\left(d\psi +\cos{\theta}d\phi\right)^2
+\f{a^2}{4}\left(d\theta^2+\sin^2{\theta}d\phi^2\right)
\eeqa

\noi
Note that the $R^2$ shrinks to a point as $u \rightarrow 0$. As $r \rightarrow \infty$ the constant $r$ hypersurfaces are given by $S^3/Z_2$. This is due to the fact that the periodicity of $\psi$ here is $2\pi$ instead of the usual periodicity $4\pi$ that gives $S^3$. There exists a two-form that is given by, 

\beqal{omega2}
\omega_2&=&\f{a^2}{2\pi}d\left(\f{\s_z}{r^2}\right)\non
&=& \f{a^2}{2\pi r^3}dr\w d\psi+\f{a^2}{2\pi r^3}\cos{\theta}dr\w d\phi+
\f{a^2}{4\pi r^2}\sin{\theta}d\theta\w d\phi
\eeqa

It can be checked that this two form satisfies (\ref{conv}).

\end{document}